# The Žďár nad Sázavou meteorite fall: Fireball trajectory, photometry, dynamics, fragmentation, orbit, and meteorite recovery

Pavel Spurný[1*], Jiří Borovička[1], Lukáš Shrbený[1]

[1]*Astronomical Institute of the Czech Academy of Sciences, Fričova 298, 25165 Ondřejov Observatory, Czech Republic*

[*]Corresponding author. E-mail: pavel.spurny@asu.cas.cz

## Abstract

We report a comprehensive analysis of the instrumentally observed meteorite fall Žďár nad Sázavou, which occurred in the Czech Republic on 9 December 2014 at 16:16:45-54 UT. The original meteoroid with an estimated initial mass of 150 kg entered the atmosphere with a speed of 21.89 km s$^{-1}$ and began a luminous trajectory at an altitude of 98.06 km. At the maximum, it reached -15.26 absolute magnitude and terminated after an 9.16 s and 170.5 km long flight at an altitude of 24.71 km with a speed of 4.8 km/s. The average slope of the atmospheric trajectory to the Earth's surface was only 25.66°. Before its collision with Earth, the initial meteoroid orbited the Sun on a moderately eccentric orbit with perihelion near Venus orbit, aphelion in the outer main belt, and low inclination. During the atmospheric entry, the meteoroid severely fragmented at a very low dynamic pressure 0.016 MPa and further multiple fragmentations occurred at 1.4 – 2.5 MPa. Based on our analysis, so far three small meteorites classified as L3.9 ordinary chondrites totaling 87 g have been found almost exactly in the locations predicted for a given mass. Because of very high quality of photographic and radiometric records, taken by the dedicated instruments of the Czech part of the European Fireball Network, Žďár nad Sázavou belongs to the most reliably, accurately, and thoroughly described meteorite falls in history.

## Introduction

Meteorite falls are products of an interaction of larger debris of asteroids with Earth's atmosphere, so their observations can tell us much about their parent bodies. Through fireball observations and subsequent meteorite recoveries, we can get direct information about internal composition and basic physical properties of asteroids and possibly comets. We can better understand the processes connected with atmospheric flight of centimeter to meter sized interplanetary bodies. Based on these observations and their analyses we can predict and describe more dangerous collisions of much larger bodies, which could cause large-scale catastrophes. Therefore, every new meteorite-producing fireball with precise atmospheric and orbital data gives us invaluable information not only about each particular event but also about its parent body. A detailed inventory of instrumentally documented falls is given in the review work Borovička et al. (2015) or in Granvik and Brown (2018). From the list of presented cases, it can be seen that meteorites were observed to fall from meteoroids of a wide range of masses, causing fireballs different by orders of magnitude in terms of energy and brightness. At the lower end, there were meteoroids of initial masses of only a few dozens of kg causing fireballs of absolute magnitude of about -10 or even slightly less, such as Bunburra

Rockhole (Bland et al., 2009, Spurný et al., 2012) or Mason Gully (Spurný et al., 2011). On the opposite end some meteorite falls were produced by large (>meter-sized) meteoroids associated with superbolide events which occur globally approximately every two weeks (Brown et al., 2002, 2013) and only very rarely were reliably documented. These cases include Tagish Lake (Brown et al., 2000), Almahata Sitta (Jenniskens et al., 2009) and the largest ever instrumentally observed bolide Chelyabinsk (Borovička et al., 2013, Brown et al., 2013, and Popova et al., 2013). In such cases, when good dynamic and photometric data are available we can obtain insight into the internal structure of the initial meteoroid for comparison with the physical structure of asteroids as determined from other kind of observations. Here we bring one of the best documented and described meteorite fall in history, the Žďár nad Sázavou meteorite fall occurred over Czech Republic on 9 December 2014 and was recorded by various instruments of the Czech part of the European Fireball Network (EN).

## Data

The Žďár nad Sázavou (shortly Žďár) instrumentally documented meteorite fall (named after a county town lying nearby the area where meteorites were found), was observed on 9 December 2014 over the Czech Republic.

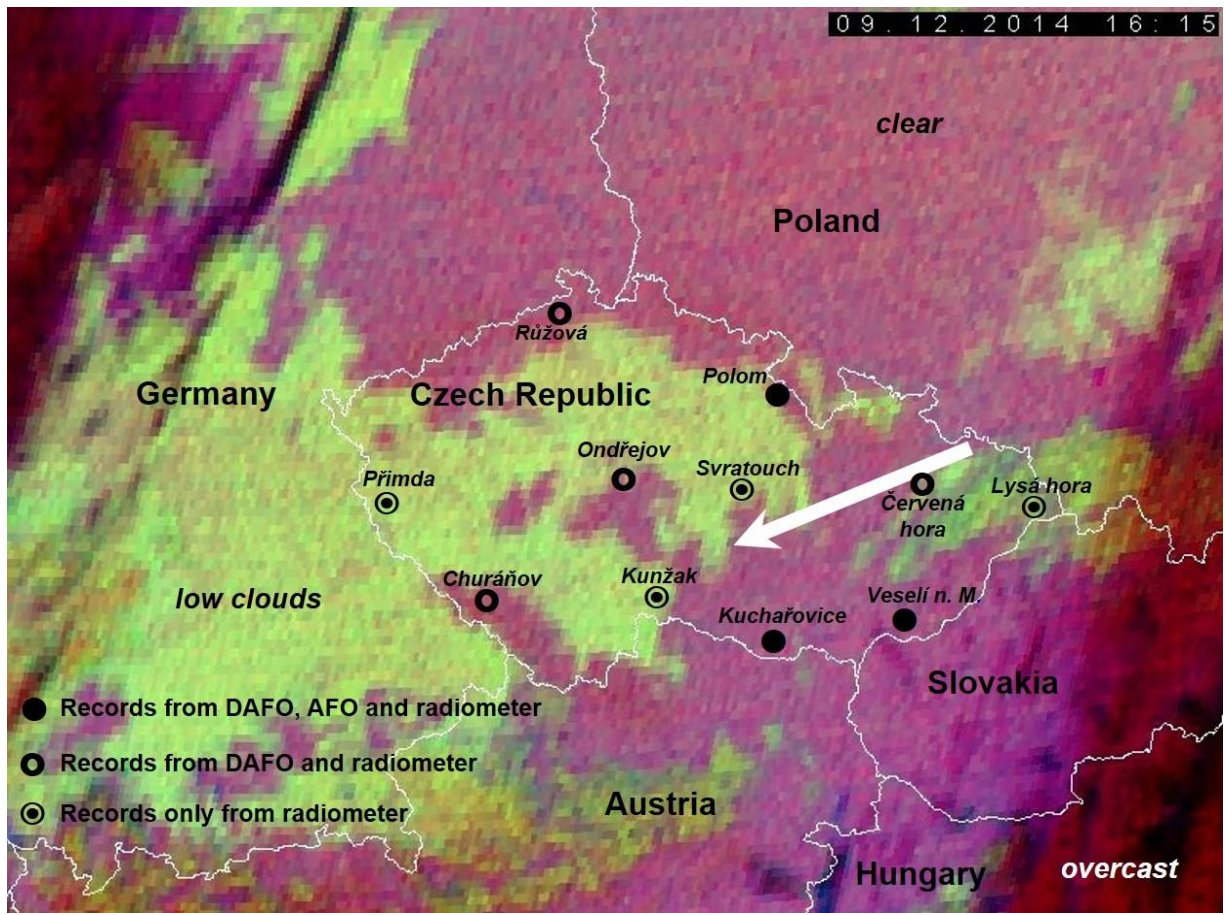

Fig. 1. Weather situation over Central Europe during the Žďár nad Sázavou bolide and distribution of Czech fireball stations (status as end of 2014) which recorded it by different instruments. Projection of the bolide trajectory is represented by a white arrow. (Source of the background image: CHMI and Eumetsat)

It occurred in the early evening, still during late local twilight, on $16^h16^m45^s$ UT and started over northeastern part of the Czech Republic close to the border with Poland. After 9.2 seconds long flight, it terminated over the Highlands County in the central part of the Czech Republic. At the maximum, it reached -15.3 absolute magnitude and riveted attention of thousands of casual witnesses not only in the Czech Republic but also practically in whole Central Europe where it was clear sky during its passage. Fortunately, after several days of cloudy skies, the weather cleared over significant part of the Czech Republic as can be seen in Fig. 1. Therefore, this extraordinary bolide could be recorded by the autonomous cameras of the Czech part of the EN. Readiness of the Czech Fireball Network (CFN) and favorable distribution of the stations in respect to its atmospheric trajectory proved to be crucial for full and detailed description of this event (see Fig 1). This was thus another tangible result of the systematic operation and modernization of the CFN. This network has been modernized several times (Spurný et al., 2007, 2017) and the last significant improvement has been realized during the two years before this event when a high-resolution Digital Autonomous Fireball Observatory (DAFO) was developed and gradually installed alongside the older "analog" (using photographic films) Autonomous Fireball Observatory (AFO) all-sky system on all Czech stations. The first stage of this difficult process terminated just before the end of 2014. Description of both used observing systems AFO and DAFO is in Spurný et al. (2017). There are several important advantages of the digital system in comparison with the original analog system. All data are immediately available and images are much simply and reliably reducible because stars are point-like and visible also near the horizon. Moreover, reduction constants can be simply transferable from other images taken by the same camera under better observing conditions. Images from DAFO, which is more sensitive than AFO, contain also more information especially in the beginning and terminal parts of the luminous trajectory. Another important advantage is the ability to take usable photographic records also during periods when it is not completely dark (twilight periods) or not completely clear. All these advantages were important also for the data acquisition and their correct analysis of the Žďár bolide because for most of stations this bolide occurred during twilight, at some of them also on a partly cloudy sky and for the distant stations in a large zenith distances (especially the western ones). Therefore, at majority of stations only digital system was in operation and took the most important records. The older AFO system recorded this fireball from three eastern stations because of later dusk (just after the nautical twilight) and these images could be used for the fully-fledged analysis also because we remotely terminated the exposure after first 6 hours (the planned exposure was 13h05m). The reason was the large phase of the Moon (only 3 days after full Moon) that would overexposed the image and the trail of the fireball would be hardly visible and measurable. It was critical especially for station Polom, which is placed north of the bolide trajectory – see Fig. 2 left image where the beginning of the bolide is very close to the Moon trail. Comparison of the all-sky images including the Žďár bolide taken at two stations where both systems were in operation is shown in Figs. 2 and 3. It is symbolic that the first stage of modernization of Czech fireball network was finished by the installation of the DAFO on the last station just in the afternoon on 9 December 2014 and that this station was the closest to the end of the fireball trajectory where it was clear. This record was the most important for the localization of the impact area and determination of the velocity and deceleration near the end of the luminous trajectory (Fig. 3).

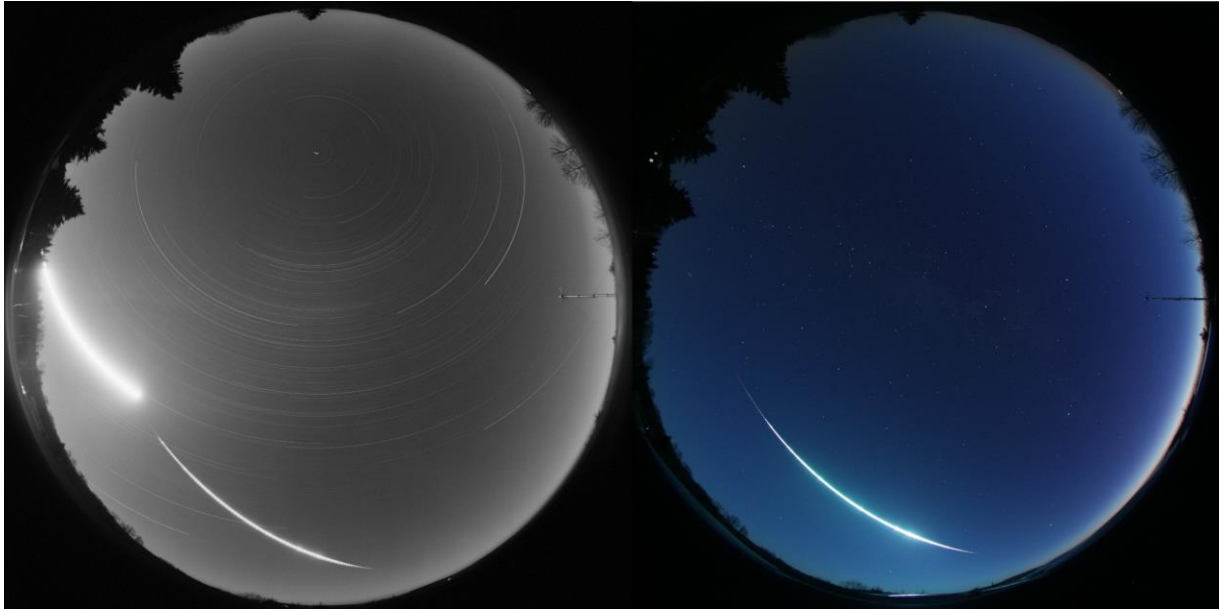

Fig. 2. Comparison of the all-sky images containing the Žďár nad Sázavou bolide taken by the AFO (left) and DAFO (right) at the station Polom.

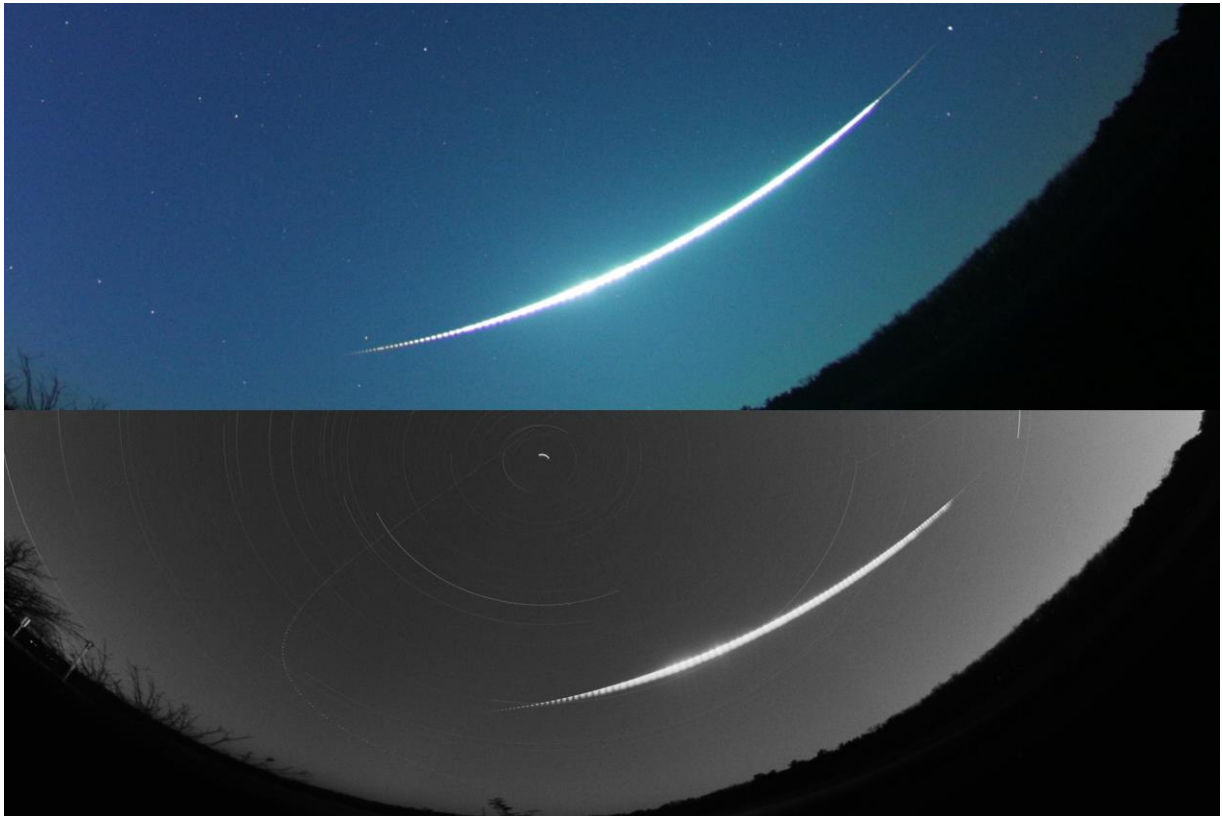

Fig. 3. Comparison of images (cut-outs from all-sky images) containing the Žďár nad Sázavou bolide taken by the AFO (bottom) and DAFO (top) at the station Kuchařovice.

Altogether, this bolide was photographed by 10 autonomous fireball cameras of the CFN; seven images were taken by DAFOs and only three (the most eastern stations where it was already sufficiently dark) by AFOs (see Fig. 1). In addition to the direct imaging, all AFOs and DAFOs in the network recorded the light curve of this bolide by their radiometers, which are integral parts of all our cameras and which are working continuously regardless of weather conditions with very high time resolution of 5000 samples per second (see Section Light curve and photometry). Along with these data, the closest AFO to the terminal part of the trajectory at the station Svratouch, where it was unfortunately overcast during fireball passage, recorded strong detonations of the bolide by its microphone. Apart from data from our instruments, this bolide was recorded also by many seismic stations in Central Europe. Altogether, this bolide became one of the best-documented cases in decades-long history of the European Fireball Network, the longest lasting continuously operational fireball network in the world.

## Trajectory

The fireball trajectory was computed using all ten available all-sky images at seven different stations. Digital images were used at all stations and as explained above, at stations Veselí nad Moravou, Polom, and Kuchařovice also images taken on the film were used.

The trajectory was first computed by the straight least-squares method of Borovička (1990). This method assumes that the trajectory is straight and it is computed by minimizing the distances in space of lines of sight from a straight line. Lines of sight were obtained by measuring up to 100 points along the meteor track on individual images. Astrometric solution was obtained by the all-sky method of Borovička et al. (1995). On digital images 200 – 400 positional stars were measured. On long exposure film images 50 – 100 reference positional points were available, consisting from the beginning and ends of star trails and a few star trails positions near the main meridian.

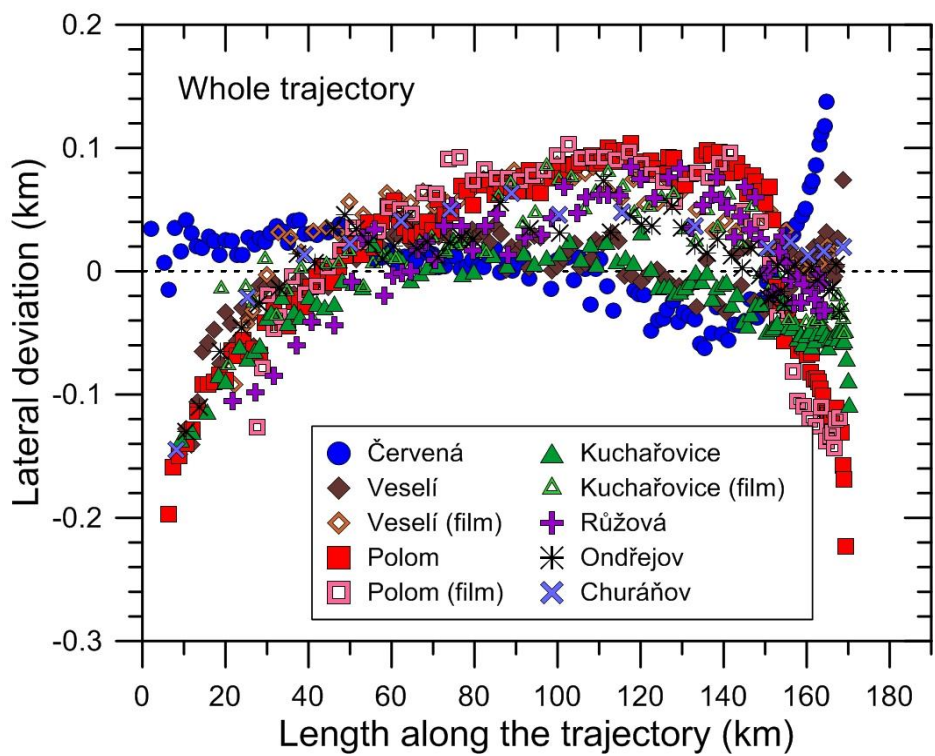

**Fig. 4.** Deviations of lines of sight at individual cameras from the average straight trajectory of the whole fireball. Total deviations are shown and the sign is positive if the line of sight crosses vertical plane above the trajectory.

The straight line method did not provide a perfect trajectory solution. Figure 4 shows the deviations of lines of sight from the solution. The sign is positive if the line of sight crosses vertical plane above the trajectory. Considering the total length of the trajectory (170 km), the absolute values of the deviations are not large. They do not exceed 200 meters in most cases. However, the deviations are not random and suggest that the trajectory was in fact curved.

The trajectory was rather shallow with the slope to the horizontal plane of about 25 degrees. Since the fireball lasted for slightly more than nine seconds, trajectory bending can be expected due to Earth's gravity. A simple estimate gives the expected deviation from the initial trajectory, $\Delta$, after elapsed time, $t$, to be $\Delta = 1/2\ g\ t^2 \cos \gamma$, i.e. about 370 meters after 9 seconds. Here $g$ is the acceleration due to gravity and $\gamma$ is trajectory slope. The curvature seen in the first half of the trajectory in Fig. 4 can be attributed to gravity. Note that station Červená lay almost directly below the trajectory, so any deviation in the vertical plane cannot be seen from it. However, there is a sharp turn in trajectory direction, visible even at station Červená, toward the end of the trajectory. This turn cannot be explained by gravity and we suspect that it was a real change in direction of flight after fragmentation of the meteoroid. The fireball was therefore divided into two parts and their trajectories were computed separately. The resulting deviations are shown in Fig. 5.

On the upper plot of Fig. 5 we can see that the first part, 143 km long, is curved even if considered separately. The scatter of points and small differences between individual cameras can be attributed to difficulties of measurements. The fireball was very bright and overexposed over large portions of the trajectory. The apparent (instrumental) width of the fireball track, when projected to the fireball distance, was about 1 km. Measurement errors of the order of 50 m are therefore quite understandable.

The second part of the trajectory was only 27 km long and no curvature was visible within the precision of measurements. When computed separately from the first part, all cameras fit nicely together (see Fig. 5 bottom).

Figure 6 compares the observed curvature of the first part of the trajectory with bending expected to be caused by gravity. Vertical deviations, i.e. deviations of lines of sight from the linear trajectory projected into the vertical plane containing the trajectory, are shown on the vertical axis. The size of the symbol is proportional to the significance of the observation, which is assumed to be proportional to $\sin \theta / R$, where $\theta$ is the angle between the line of sight and the fireball vertical plane and $R$ is the range of the fireball (to the point on the trajectory closest to the line of sight). Data from station Červená hora have near zero vertical deviations but zero significance. Stations which captured the fireball from a side must be used to evaluate vertical deviations.

The thick grey line shows one possible course of deviations caused by gravity. To compute them, relative time must be assigned to each length. The shutter breaks on the fireball images were used for that. The considered part of the trajectory was traveled by the fireball in 6.7 seconds. The initial velocity of 21.89 km s$^{-1}$ decreased to 19 km s$^{-1}$ toward the end. The dynamics of the fireball is discussed in more detail in sections Velocity and Orbit, and Fragmentation model.

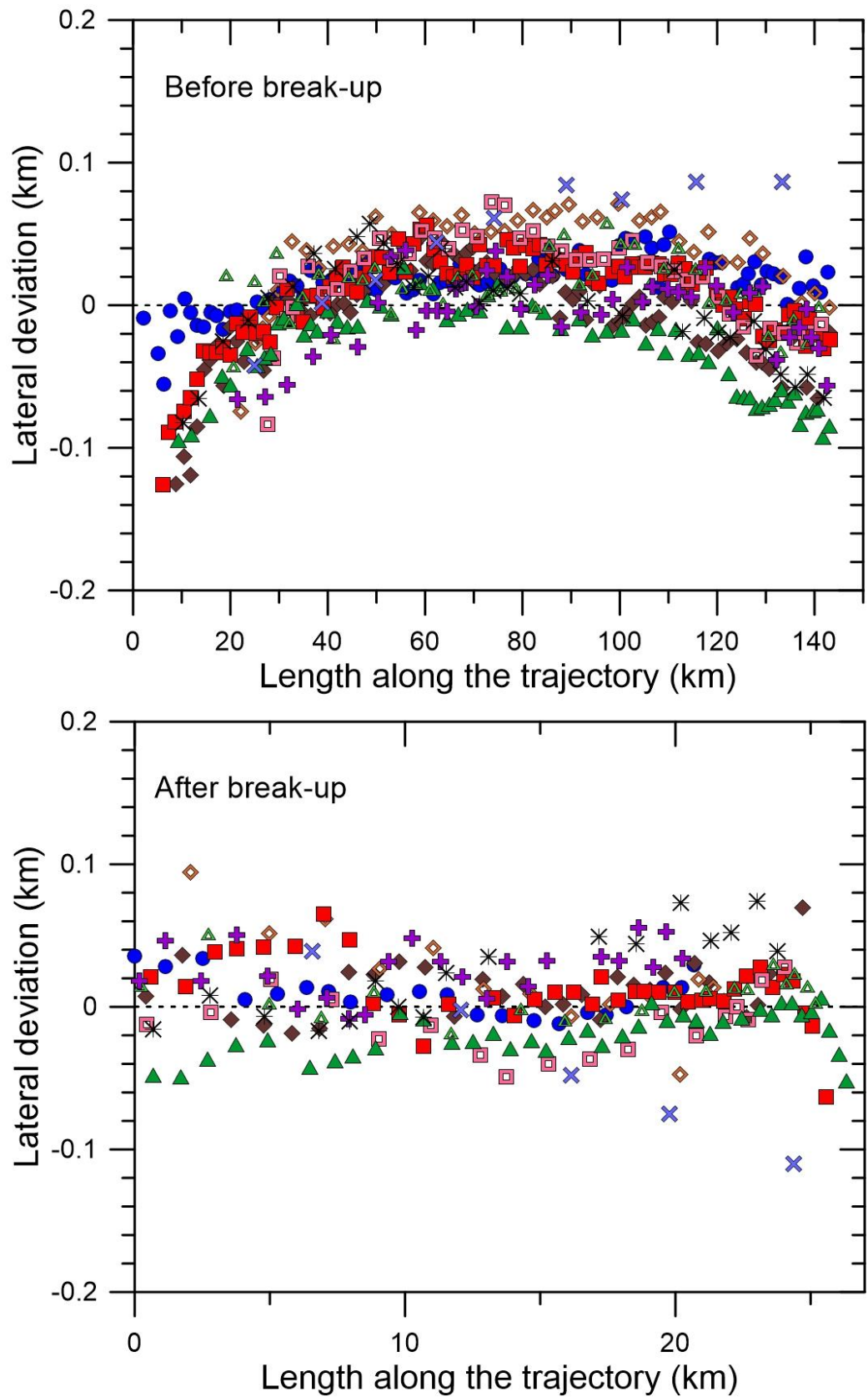

**Fig. 5.** Deviations of lines of sight at individual cameras from the average straight trajectory computed separately for the first 143 km of fireball trajectory, i.e. until the break-up (upper panel), and the rest of the fireball. For the legend, see Fig. 4.

The observed and expected trajectory curvature agrees well. Only at the beginning, the measured points deviate more but they have low significance. There is some freedom how to set the gravity bent path within the observations. The adjustment was made so that the end point corresponds to the independently computed beginning point of the second part of the trajectory.

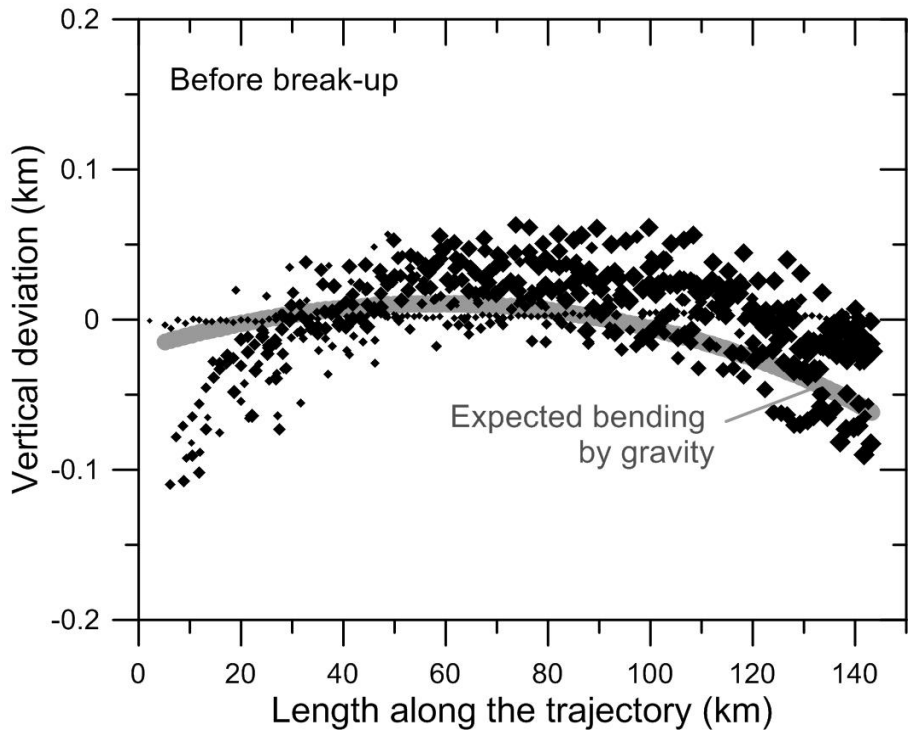

**Fig. 6.** Vertical deviations of lines of sight from the average straight trajectory for the first part of the fireball. The sizes of the symbols are proportional to the significance of the points for the vertical deviation measurements. The significance depends on the distance to fireball and the angle between the line of sight and the vertical plane. The thick grey curve is the solution for gravity bent trajectory.

The change of the trajectory after the fragmentation is visualized in Fig. 7. The left part (panels a, b) shows the deviation of position of the body and the right part (panels c, d) the difference in the direction of flight, both relative to the straight trajectory solution for the main part before fragmentation. Before fragmentation, the trajectory was bent by the gravity, so vertical deviation and difference of radiant zenith distance are smooth functions of length along trajectory. The horizontal deviation and difference in azimuth were zero because the bent trajectory was assumed to remain confined in the same vertical plane as the linear solution trajectory. At the fragmentation, the direction of flight changed. The emerging fragment turned down and to the left from the vertical plane in the direction of the flight. The difference in azimuth was $da = 0.66$ degrees and the difference in zenith distance was $dz = 0.19$ degrees. The overall change of the direction of flight was $du = \sqrt{((da \sin z)^2 + (dz)^2)} = 0.63$ degrees. At the velocity of 19 km s$^{-1}$, this angle means that the fragment gained a lateral velocity of 200 m s$^{-1}$. Note that the measured fragment was the only one and quite dominant body visible on the all-sky images. The fragmentation is apparent only by this slight change of direction on the photographs. At the end of the fireball, the main fragment deviated about 250 m in horizontal direction (southwards) and more than 100 m in vertical direction (downwards) from the point extrapolated from the original pre-fragmentation trajectory (Fig.

7a, b). Due to increasing deceleration, the fragment trajectory was expected to be bent by gravity by about 0.25 degrees (Fig. 7c) but the actual deviation in meters from the linear fragment-trajectory solution was too small to be observed.

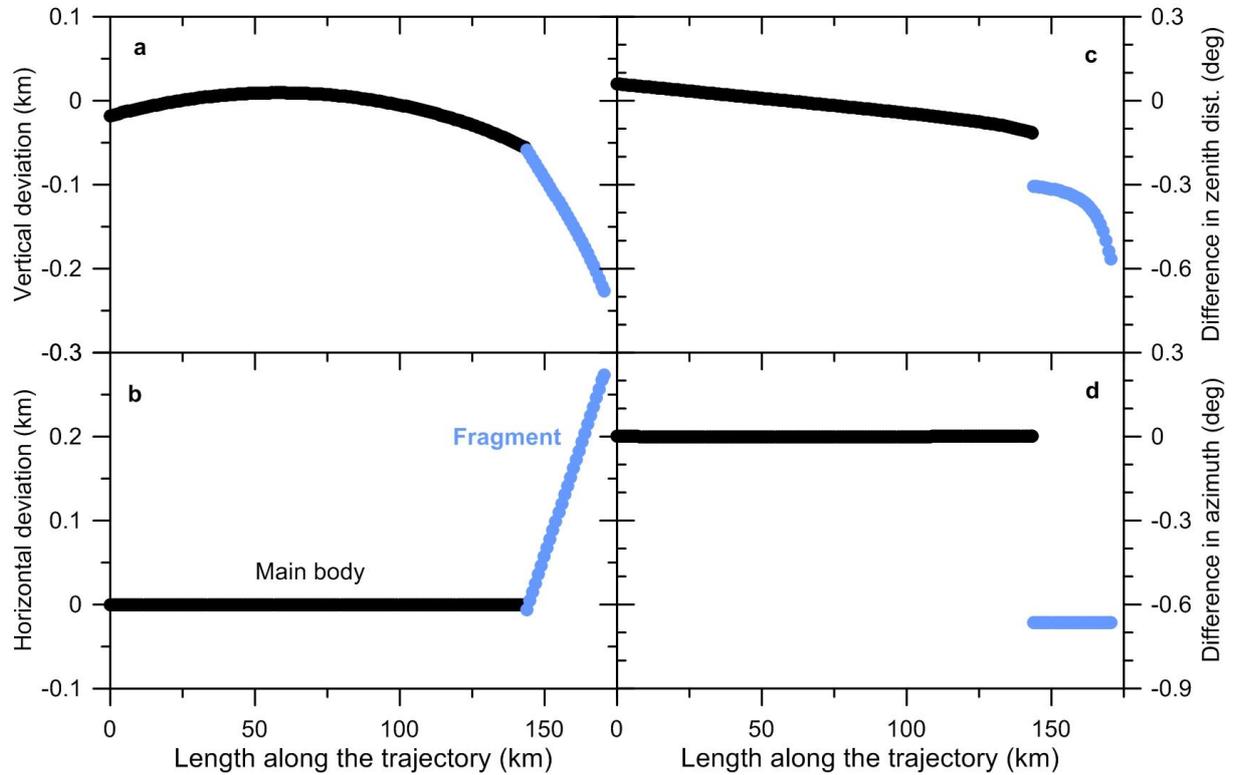

**Figure 7.** The differences in position and radiant of the fireball relatively to the linear trajectory solution for the main body before the break-up. The main body differs from the linear solution only due to gravity. The fragment, in addition, changed the direction of flight at the break-up point.

**Table 1.** Geographical coordinates and apparent radiants for two independently computed parts of the fireball trajectory.

|  | Longitude °E | Latitude °N | Height km | RA ° | Decl. ° | Azimuth ° | Zenith dist. ° |
|---|---|---|---|---|---|---|---|
| Beginning | 18.00129 | 49.94052 | 98.062 | 65.241 | 30.440 | 252.833 | 63.965 |
|  | ±4 | ±6 | ±4 | ±4 | ±4 | ±4 | ±4 |
| End of main trajectory | 16.31110 | 49.58873 | 36.310 | 65.088 | 30.569 | 251.542 | 64.926 |
|  | ±4 | ±6 | ±4 | ±4 | ±4 | ±4 | ±4 |
| Beginning of fragment | 16.30344 | 49.58711 | 36.035 | 65.451 | 31.129 | 250.866 | 64.751 |
|  | ±4 | ±7 | ±5 | ±23 | ±20 | ±22 | ±20 |
| End | 15.99090 | 49.51609 | 24.710 | 65.232 | 31.311 | 250.628 | 64.705 |
|  | ±4 | ±6 | ±4 | ±23 | ±20 | ±22 | ±20 |

Table 1 gives the geographical coordinates (in the WGS84 system) of the observed beginning and end points of the main part of the trajectory until the fragmentation and the final part after the fragmentation. Naturally, both parts should have a common point – the fragmentation point. However, the exact location of the fragmentation point is not apparent on the images. Both parts of the trajectory were computed independently from different sets of measurements.

We can just say that the fragmentation and the change of flight direction occurred between heights 36.0 and 36.3 km.

Table 1 contains also equatorial and azimuthal coordinates of the apparent radiant. Equatorial coordinates have not been converted to a standard equinox, they are valid for the day of the fireball. The changes along the trajectory have been caused by the gravity bending and by the change of flight direction after the fragmentation. Azimuths and zenith distances of the radiant were computed from the equatorial coordinates for the given geographical point. The changes along the trajectory are due also to the Earth's curvature (i.e. azimuths and zenith distances would change even if right ascension and declination were constant along the trajectory). Azimuths are counted positively from the south to the west. The standard deviations listed are formal standard deviations of the straight trajectory solution.

Projection of the fireball trajectory on the map of Czech Republic is shown in Fig. 1. The fireball was first photographed at the height of 98 km, about 20 km to the north-west from Ostrava, not far from Czech-Polish border. The body headed toward west-south-west and passed almost overhead (only 5.5 degrees from zenith) the station Červená hora when it was at height 80 km (geographical coordinates of all stations are given in the supplementary file Supplementary_data.xlsx). After the break-up at the height of 36 km the largest fragment continued on slightly changed trajectory and ceased to be visible at the height 24.7 km about 30 km from Jihlava town and 6 km from Žďár nad Sázavou town. The trajectory was 170 km long, from what 143 km was before the break up and 27 km after the break-up.

Stations Veselí nad Moravou and Kuchařovice had a nice side view of the fireball from the south and station Polom from the north. The data from these three stations and Červená hora defined the trajectory quite well, including the gravity bending and the change after break-up. Červená hora was ideal for measurement of the velocity at the beginning, while Kuchařovice were closest to the end. Unfortunately, station Svratouch, which was much closer, was clouded out. The three western stations were more distant from the fireball and had poorer geometry – especially Churáňov, where the fireball was almost stationary. These two stations got low weights in our calculations. Nevertheless, as it can be seen in Fig. 5, their data are fully consistent with the trajectory solution.

## Velocity and orbit

Fireball velocity along the trajectory was measured using the time marks on fireball images produced by shutters inside cameras. Digital cameras are equipped by LCD shutters placed just behind the lenses. The shutters alternate between opaque and transparent states with the frequency of 16 Hz. They are controlled by the precise GPS PPS (Global Positioning System pulse per second) time signal. After the start of each whole second, one opaque state is skipped. As a result, the fireball image starts with one long dash at the beginning of second, followed by 14 short dashes. The measurements of the leading edges of the dashed provide the positions (lengths) along the trajectory as a function of time. The situation is illustrated in Fig. 8. All measurements were done manually on the computer screen. Absolute timing of long dashes was provided by radiometric curves (see below).

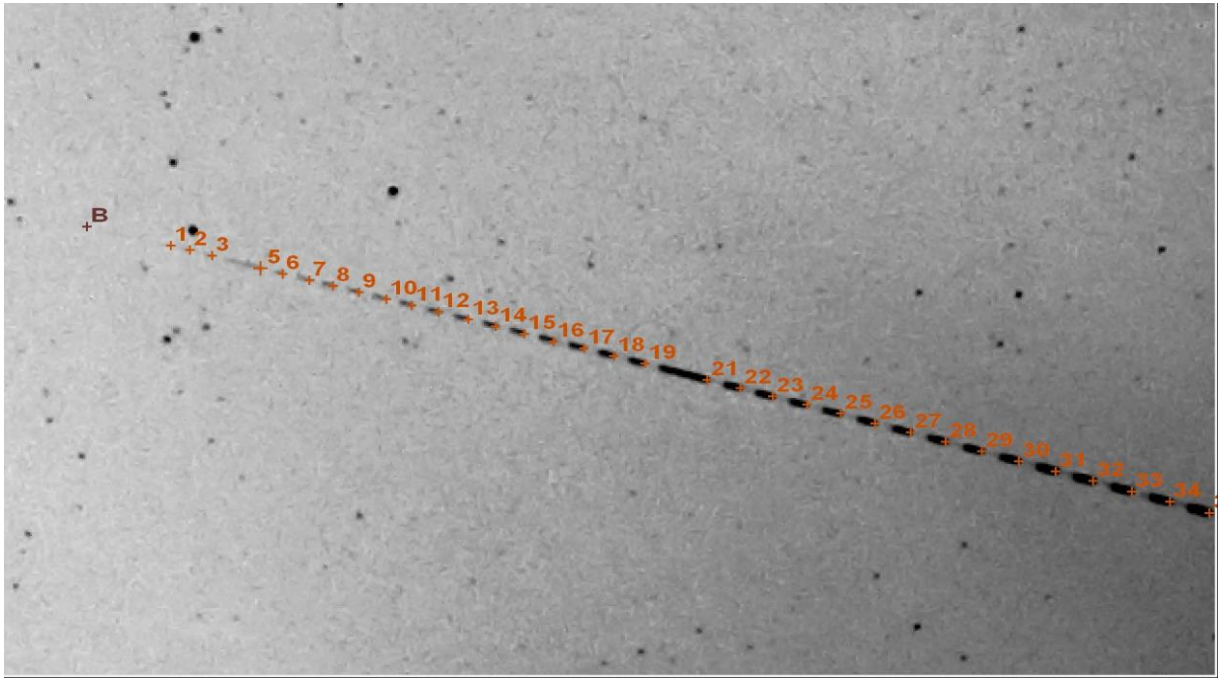

**Figure 8.** Beginning of the fireball as photographed by the digital all-sky camera at station Červená hora. The image was inverted and converted to grayscale; contrast was enhanced. The numbered orange crosses are measurements of shutter breaks. The interval between shutter breaks is 1/16 s. One shutter break is skipped at the beginning of each second.

The film cameras use mechanical rotating three-arm shutter located near the focal plane just above the film. The shutter frequency is 15 Hz. There is no absolute timing. The adjustment of relative time to absolute time was obtained using the time-length dependency from digital cameras.

In total 682 measurements of length as a function of time was obtained on 8 cameras. They are provided in the supplementary file Supplementary_data.xlsx. From these data velocity and deceleration was studied. As a preliminary solution, the data were fitted by the single-body 4-parameter fit of Ceplecha et al. (1993), separately for the main body before the break-up and for the fragment after the break-up. The corresponding trajectory solution was used for each fit. Final solution was obtained by detailed fragmentation modeling, which considered not only the dynamic data but also the light curve. The light curve revealed that there were more than one fragmentation event. Details are given in the corresponding section.

Fragmentation modeling yielded the initial velocity $21.89 \pm 0.02$ km s$^{-1}$ (the 4-parameter fit gave 21.886 km s$^{-1}$ with formal error 3.2 m s$^{-1}$). Figure 9 shows the distribution of measurements of the main body. Time is on horizontal axis. Vertical axis shows the difference of measured length along the trajectory from the length expected for the given time and constant velocity of 21.89 km s$^{-1}$. Except for some outliers, all measured points follow the same trend. The velocity was constant for about 3 seconds and then the lag started to increase due to deceleration. Due to geometric conditions (low angular speed), velocity could not be measured at stations Ondřejov and Churáňov. Data from distant station Růžová are relatively poor. Similarly, Kuchařovice, especially the digital camera, gave poorer data at the beginning due to unfavorable geometric conditions (large distance and closeness of the fireball to the radiant). The digital Kuchařovice data were, nevertheless, crucial for measuring the velocity at the end of the fireball (after the break-up). Overall, taking into account the difficulties connected with fireball brightness, the agreement between various independent cameras is good.

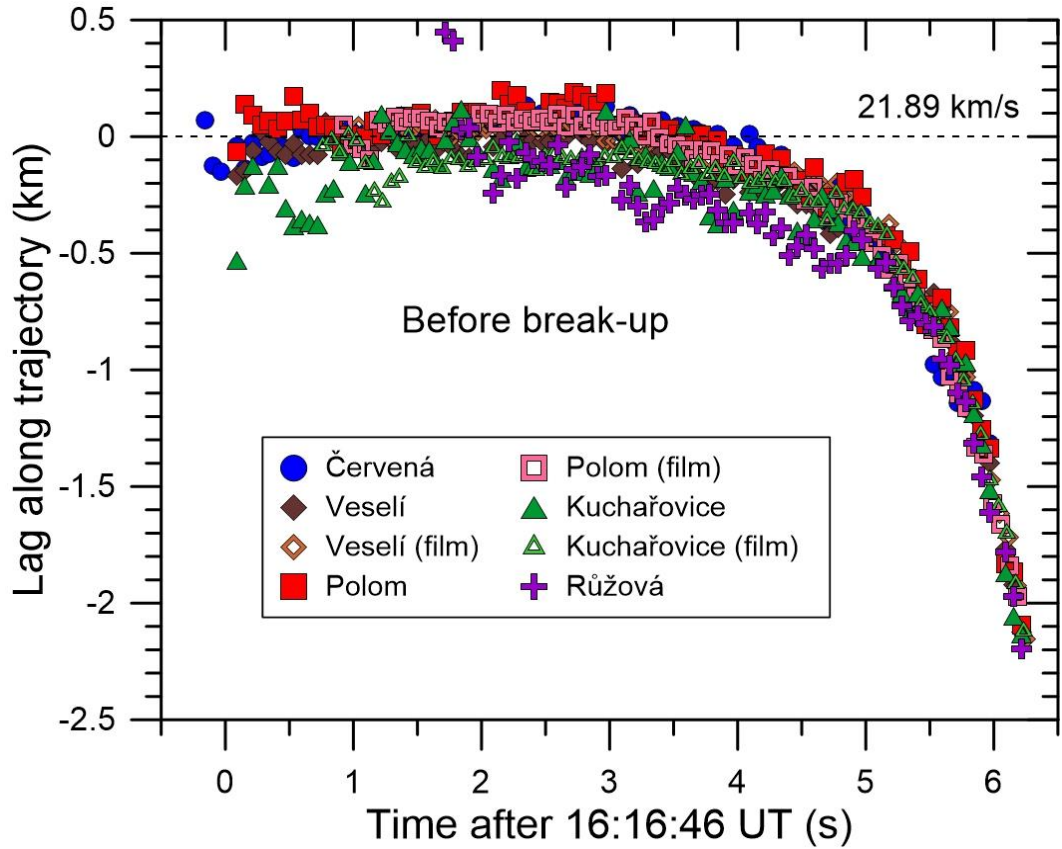

**Fig. 9.** Measured difference between the position in trajectory and the position expected for constant velocity of 21.89 km s$^{-1}$ as a function of time. This kind of graph enables to compare velocity measurements at different station. Only the data before the geometrically observed fragmentation at height 36 km are given.

Before the break-up, the velocity decreased to 19.2 km s$^{-1}$. After the break-up, the deceleration was strong and the velocity further decreased quickly. Fig. 10 shows the velocity as a function of time according to the dynamic fits and the fragmentation model. Both functions differ slightly only near the break-up point and at the very end. For the last measurement at time 16:16:54.72 s (height 24.9 km), fragmentation model gives velocity 4.8 km s$^{-1}$ while the 4-parameter fit gives 4.9 km s$^{-1}$. The strongest deceleration (–7.6 km s$^{-2}$) occurred around time 6.9 s (height 31.5 km).

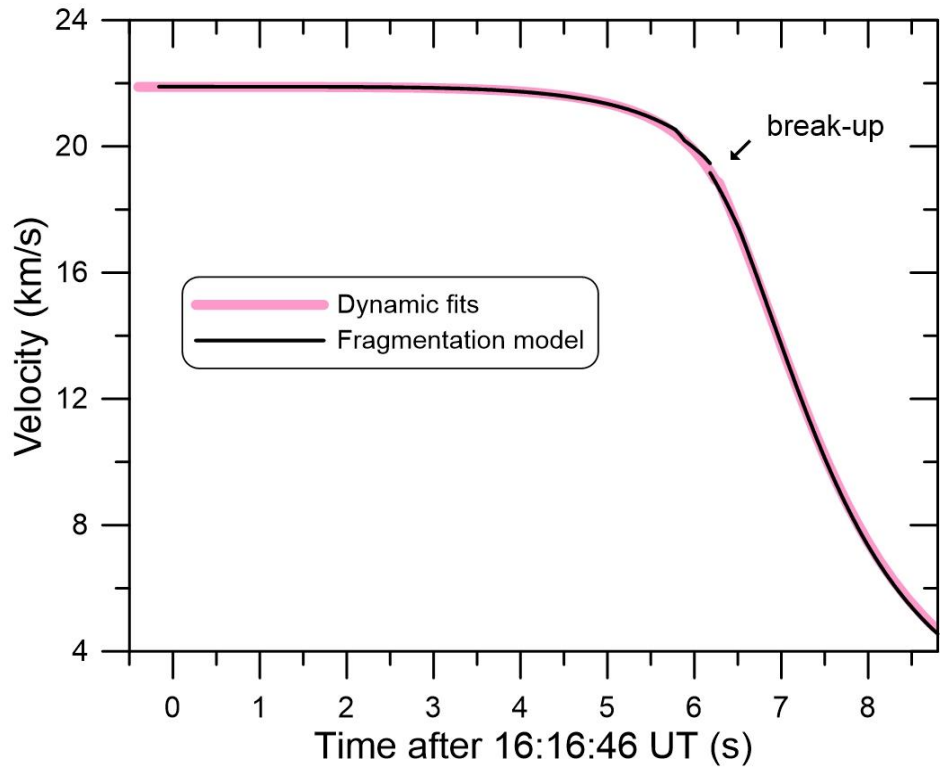

**Fig. 10.** Velocity as a function of time derived from data in Fig. 9 (and similar data after the break-up) using two approaches. Dynamic fits use the single body theory for shutter break measurements, separately before and after the break-up. Fragmentation model contains more fragmentation points and explains also the light curve.

**Table 2.** Fireball radiant and heliocentric orbit. Apparent radiant is valid for the beginning point as listed in Table 1.

| | |
|---|---|
| *Apparent radiant (J2000.0) and entry velocity* | |
| Right ascension | 64.997° ± 0.004° |
| Declination | 30.406° ± 0.004° |
| Velocity | 21.89 ± 0.02 km s$^{-1}$ |
| *Geocentric radiant (J2000.0) and velocity* | |
| Right ascension | 69.298° ± 0.010° |
| Declination | 26.963° ± 0.009° |
| Velocity | 18.56 ± 0.02 km/s |
| *Orbital elements (J2000.0)* | |
| Semimajor axis | 2.093 ± 0.006 AU |
| Eccentricity | 0.6792 ± 0.0010 |
| Perihelion distance | 0.6715 ± 0.0002 AU |
| Aphelion distance | 3.514 ± 0.012 AU |
| Inclination | 2.796° ± 0.009° |
| Argument of perihelion | 257.721° ± 0.014° |
| Ascending node | 257.262° ± 0.010° |
| Time of perihelion | 2012 Jan 13.0 ± 5 days |

From the known radiant, entry velocity, and fireball time, the pre-entry heliocentric orbit was computed by the analytical method of Ceplecha (1987). The method was slightly refined in the sense that for the correction to Earth's rotation the average coordinates and height along the trajectory was used while for the correction to zenith attraction the beginning point of the trajectory (and the apparent radiant at that point) was used. The apparent and geocentric radiant and orbital elements (all for standard equinox J2000.0) are given in Table 2. The orbit has the perihelion near Venus orbit, aphelion in the outer main belt, and low inclination. It is an orbit of obviously asteroidal origin with the Tisserand parameter relatively to Jupiter $T_J$ = 3.42 and orbital period of 3.03 years. Two small near-Earth asteroids 2011 WU74 and 2011 WV74 have orbits of the same character, though with little bit different orientation and higher inclinations (5.9° and 7.2°, respectively). The orbit of Žďár nad Sázavou and both asteroids are plotted in Fig. 11. Possible relation of these three objects may be subject of further study, although most probably the similarity of the orbits is coincidental. The dissimilarity criteria are not particularly low, the criterion of Southworth and Hawkins (1963) gives $D_{SH}$=0.13 and the criterion of Drummond (1981) gives $D_D$=0.05 between Žďár nad Sázavou and both 2011 WU74 and 2011 WV74. No known meteor shower is closer to the orbit of Žďár nad Sázavou than these two asteroids.

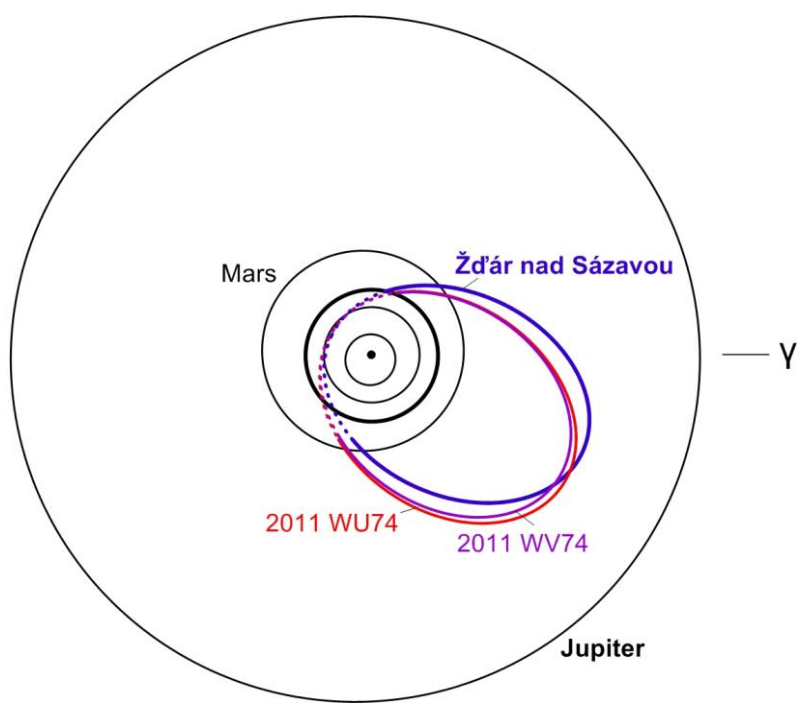

**Fig. 11.** Orbits of Žďár nad Sázavou and near-Earth asteroids 2011 WU74 and 2011 WV74 in the projection to the plane of ecliptic. The parts of orbits which lie below the ecliptic are dashed. The orbits of the asteroids were taken from the JPL database.

## Light curve and photometry

Fireball brightness was measured using digital photographs, film photographs, and radiometers. Radiometers are parts of both analog and digital cameras and provide the most detailed and reliable light curves in relative units. For converting radiometric curves into absolute magnitudes, photographic data are used.

Radiometers are sensitive photomultiplier tubes aligned vertically. They measure the overall brightness of the sky 5000 times per second. The response is linear in wide range of intensities (about $1:10^6$). There is no information about the position of the light source(s) on the sky. The background sky signal (in the absence of transient sources) is kept constant during the night using a high voltage control. Since bright Moon was expected on the sky most of the night of December 9/10, 2014, the background value was set to 8000. At the time of the fireball in early evening, nevertheless, the Moon was still below horizon. The higher background value caused little bit higher sensitivity and lower dynamic range of radiometers in comparison with our usual no-Moon value of 1000.

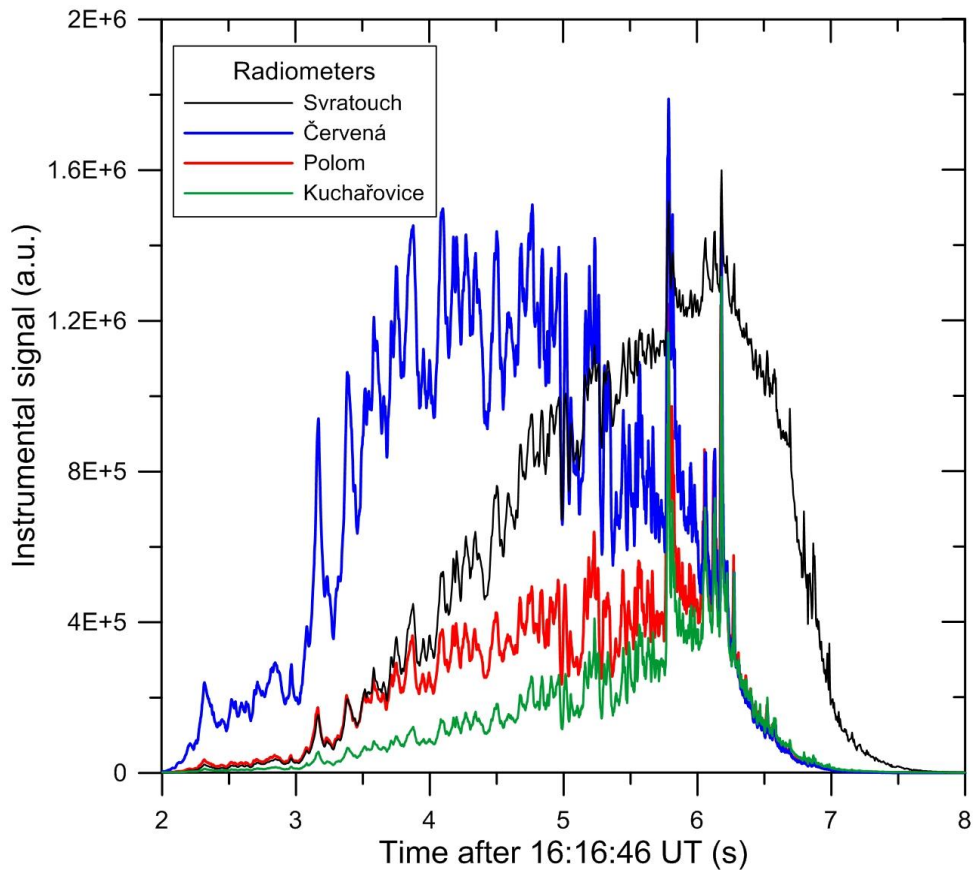

**Fig. 12.** Radiometric signals at four different stations as a function of time.

Radiometers are supplied by absolute time using GPS PPS signal. Fig. 12 shows raw radiometric signals at four stations in relative units in linear scale as a function of time. The marked differences in the overall shape of the curves are caused mostly by different geometric conditions as the fireball range varied widely from station to station and with time. In addition to the usual inverse square law, the effect is further enhanced by decreasing sensitivity of radiometers with increasing zenith distance of the source. There are also differences in sensitivities of individual radiometers. Nevertheless, the same waves and spikes are present in the light curves from all radiometers and are surely real effects. The noise was much lower (about 50 units) than the fireball signal in the bright phase.

From Fig. 12 it is obvious that the fireball was closest to station Červená hora in the first half of the trajectory and to station Svratouch in the second half (cf. Fig. 1). In fact, the signal at

Svratouch became partly saturated, i.e. entered the non-linear part of the response, from time 4.7 to 6.7 s. This is evident e.g. by the low amplitude of the main spikes in comparison with other stations. The Svratouch curve was therefore not used for further evaluation. On the other hand, it provided the best signal toward the end of the fireball (after time 6.7 s), despite the fact that the station was in fog and the fireball was not directly visible here.

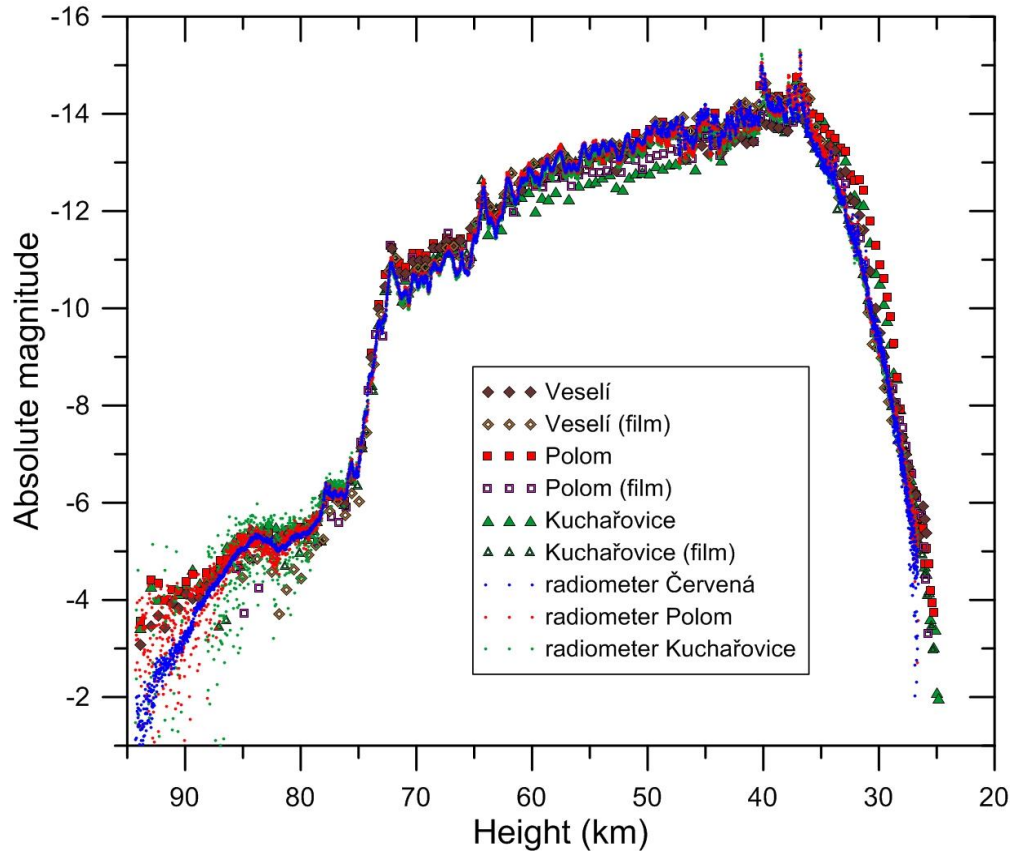

**Fig. 13.** Light curve as a function of height. Data from six cameras and three radiometers are shown. Radiometric data are presented in the resolution of 0.002 s.

After the fireball trajectory and its position as a function of time was computed, the radiometric signal could be corrected for fireball range and zenith distance. At the same time, photographic photometry of the fireball was done. On short-exposure digital images, the signal of whole shutter breaks was compared to signals of point-like stars, taking into account the difference in exposure times for the meteor (1/32 s) and stars (35 s). Atmospheric extinction was taken into account. The response of digital cameras was linear unless the signal was saturated. On long-exposure film cameras, the signals of shutter breaks when scanned across the meteor were compared with signals of star trails, taking into account different angular velocity of stars and meteor. Schwarzschild coefficient of 0.8 was assumed. The fireball was, however, so bright and so close that it saturated both digital and film cameras along most of its trajectory (basically when it was brighter than absolute magnitude –8 to –9). Photographic photometry was most reliable during a plateau on the light curve near the fireball beginning, when the magnitude reached about –5. The signal was strong enough but not yet saturated. Radiometric curves were shifted individually to match photographic photometry in this part. Note that both photographic and radiometric data were absolutely timed, so no shift along horizontal axis was needed. Subsequently, approximate saturation correction was applied to the photographic data to match the radiometric curves in the bright part of the fireball. The correction was done by changing the slope of the characteristic curve

(i.e. the relation of logarithm of the signal to magnitude) at some point. The resulting light curve as a function of height is shown in Fig. 13. The data are provided also in the supplementary file Supplementary_data.xlsx (only non-saturated data are given from photographs). Generally, there is good agreement of all data. Some camera data deviate up to about 1 magnitude in the saturated part, which is not surprising. At the beginning, radiometric data, except from the closest station Červená hora, show scatter due to low signal. The scatter could be reduced by averaging more points (data shown use 10-point averaging, i.e. the original resolution 5000 Hz was reduced to 500 Hz). Film cameras, since they are less sensitive than digital cameras, show fever data and less reliable data at the beginning and at the end of the fireball.

A striking aspect of the light curve is an enormous increase of brightness from –6.5 to –11 magnitude (i.e. almost 100×), which occurred within 0.3 s between heights 75 – 72 km. Another significant increase by 1.8 magnitudes in 0.15 s occurred between heights 65.5 – 64 km. The bright part of the light curve is shown in more detail in Fig. 14. There were many other humps during the gradual increase of brightness. The increase was slowing down when the fireball approached the height of 40 km. But at 40.3 km a very steep brightening occurred by 1.4 magnitudes in less than 0.02 s culminating in a local maximum of –15.1 mag at 40.2 km. This flare was followed by other spikes in the next half second. The fireball reached its maximum brightness of –15.26 mag (± 0.10 mag) at one of them at a height of 36.8 km. This height corresponds reasonably well with the geometrically observed fragmentation. The light curve therefore confirms that the fragment, which was observed further down, originated in this last big fragmentation event. The flare was caused by the dust released during the fragmentation. After that the bolide brightness started to fade quickly, although there were numerous spikes even on the descending part of the light curve.

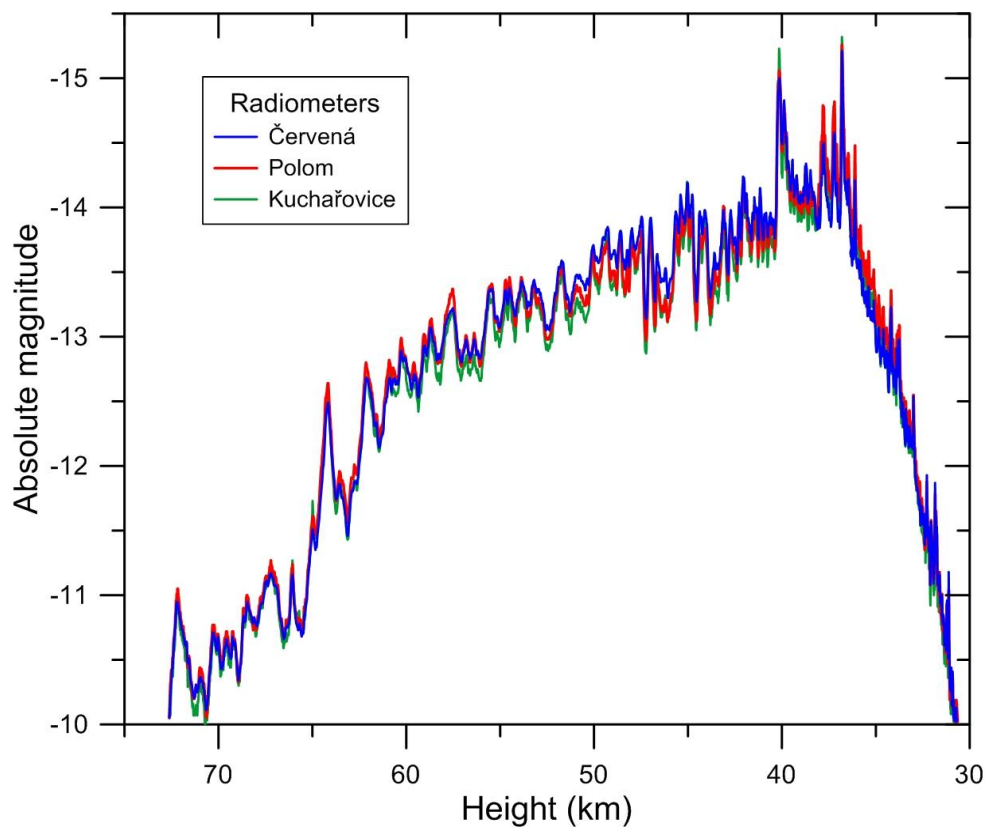

**Fig. 14.** Detail of the light curve in Fig. 13 showing only radiometric data in the bright part of the fireball.

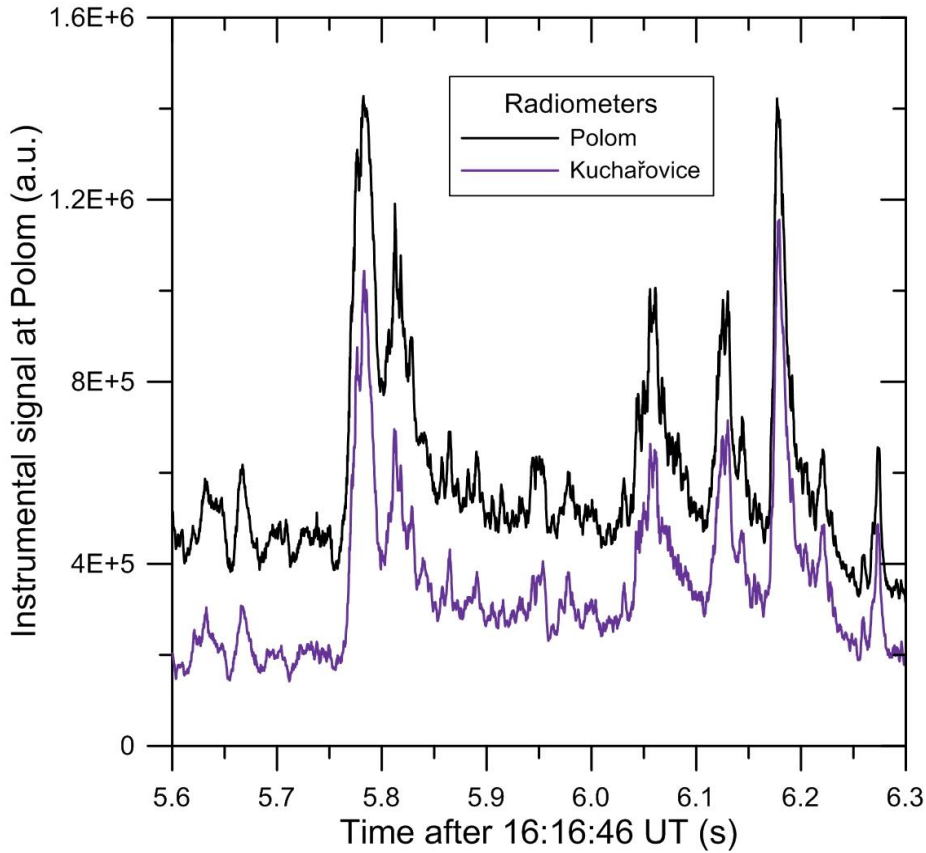

**Fig. 15.** Detail of the radiometric data at two stations for the fireball maximum. Uncalibrated data are shown in linear scale in the full resolution of 0.0002 s. The Kuchařovice data have been offset down for clarity.

The complexity of the light curve is demonstrated in Fig. 15, where two radiometric curves from two widely separated (165 km) stations are shown. Only the part of the light curve containing the brightest flares is presented. Vast majority of the minor features are present in both datasets, which nicely demonstrates perfect compliance of both records and the fact that each this feature is real as well.

The total radiated energy was $1.4 \times 10^9$ J which corresponds to 0.33 T TNT. This value was computed using the conversion factor of Ceplecha et al. (1998) giving that zero magnitude meteor has radiative output of 1500 W.

## Fragmentation model

Our semi-empirical fragmentation model was applied in order to explain the dynamic and photometric data. The principles of the model were described together with the analysis of the Košice meteorite fall (Borovička et al. 2013). Figure 16 shows the comparison of the observed and modeled light curve. The modeled light intensity at any time is proportional to the loss of kinetic energy of the meteoroid. The contributions of all fragments existing at that time are summed together. The loss of kinetic energy includes loss of mass due to ablation and loss of speed due to atmospheric drag. The luminous efficiency was assumed to be a function of speed and mass of the fragment. The speed dependence was taken from ReVelle and Ceplecha (2001). The mass dependence was taken from the same work but scaled so that the upper limit of luminous efficiency (for masses >> 1 kg) was 5% and the lower limit (for masses << 1 kg) was 2.5% at the speed of 15 km s$^{-1}$. This is identical to Košice modeling (Borovička et al., 2013).

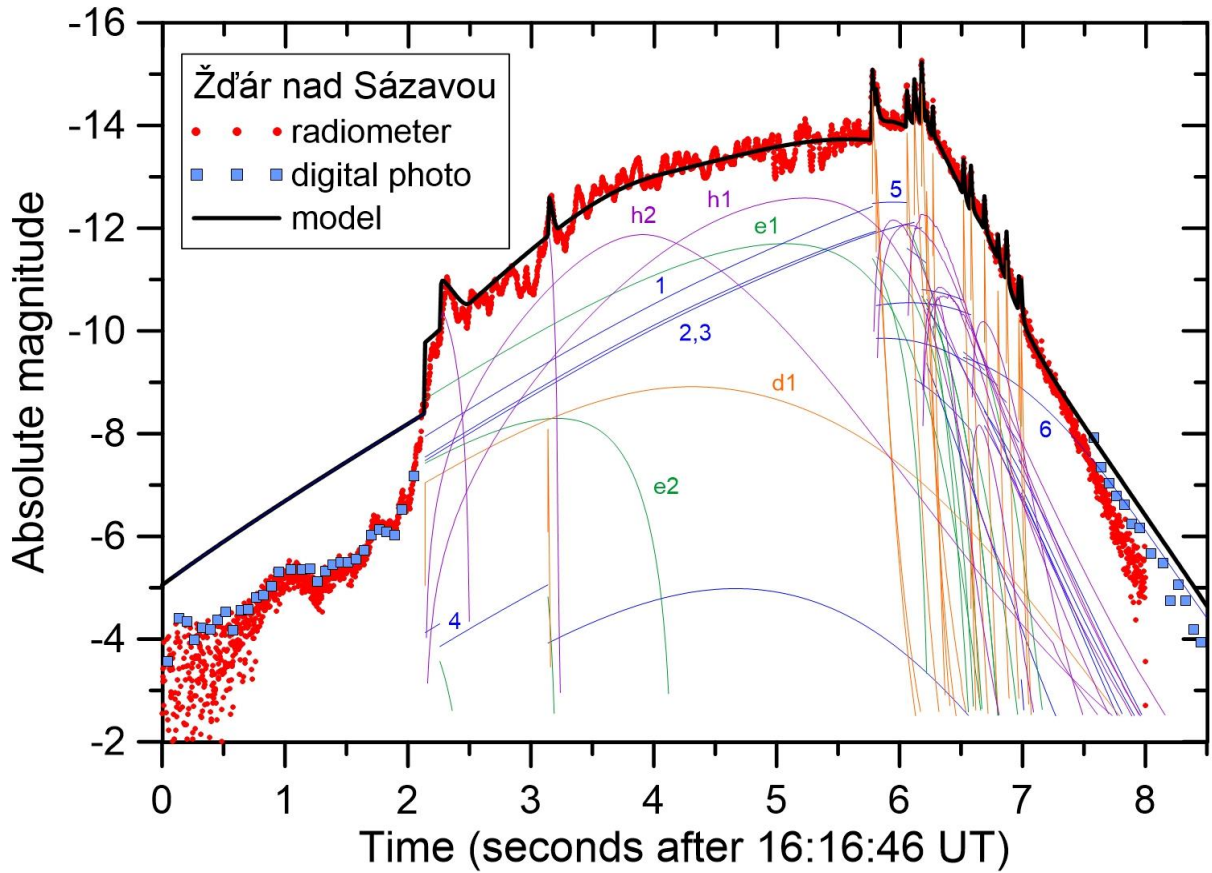

**Fig. 16.** Comparison of observed and modeled light curve. Observed data from Polom radiometer and the non-saturated part of Polom digital photograph photometry are shown. The contributors to the modeled curve are shown by thin color lines. See text for details.

To explain the whole light curve, the entry mass of the meteoroid was set to 150 kg. The first two seconds of the light curve could not be fitted well with the model. The computed brightness was too high even with the use of low ablation coefficient of 0.001 s$^2$ km$^{-2}$ (=kg/MJ) for this part. On the other hand, the general slope (brightness increase) was nearly right. We suppose that the luminous efficiency was lower at this beginning phase when the meteoroid surface was not yet fully heated. It is also possible that the meteoroid was not spherical and the cross-sectional area was lower than corresponding to sphere (the model assumed ΓA=0.8, where Γ is the drag coefficient and A is the shape factor). In any case, this beginning part is not important for the fragmentation history.

Significant fragmentation is needed to explain the dramatic rise of brightness around time 2.1 s (height 74 km). Note that the time is counted from 16:46:46 UT. The fireball beginning (Table 1, Fig. 8) was recorded at time –0.4 s. At time 2.1 s, the meteoroid was disrupted into a number of smaller fragments. The model contains three kinds of objects, which can be produced at fragmentation: regular fragments, eroding fragments, and dust. Regular fragments are characterized by their mass and are subject to just normal ablation until the next fragmentation. Unless otherwise noted, the ablation coefficient was assumed to be 0.005 s$^2$ km$^{-2}$, density 3200 kg m$^{-3}$, and ΓA=0.8 to 0.7 (lower for fragments in denser atmosphere toward the end). Dust is a group of larger number of smaller fragments released immediately (i.e. within the time resolution of the model, which was 0.01 s), each of them ablate individually. Dust particles can either have all the same mass or there can be a range of masses and a mass distribution index. Even in that case, for computational simplicity, dust is grouped into mass bins of the same mass. We used 4 bins per order of magnitude of mass (for example logarithms of masses in kg –4, –4.25, –4.5, –4.75, and –5). The mass distribution index was set to 2 in all cases. Eroding fragments are fragments, which are releasing dust gradually over time. In addition to mass loss by ablation (evaporation) they are subject to loosing mass in forms of small fragments/dust. The dust loss rate is characterized by an erosion coefficient, which has the same units as ablation coefficient but is always larger. An eroding fragment contributes by two ways to the overall light curve. The fragment itself is dominant only at the very beginning. As more and more dust is being released, the dust provides more light because of its large total cross section. Dust ablation forms a hump on the light curve. The shape and duration of the hump depends on the dust mass range, the erosion coefficient and the atmospheric density at the actual height. The ablation coefficient, $\Gamma A$, and density were assumed to be the same for eroding fragments and dust particles as for regular fragments.

In Fig. 16 the contributors to the summary light curve are shown by thin lines. Regular fragments are in blue, immediately released dust in orange, eroding fragments in green, and dust released from eroding fragments in purple. In order to explain the main characteristics of the light curve, the meteoroid was modeled to break into three major regular fragments (marked 1, 2, 3 in Fig. 16) and five eroding fragments (marked $e$1, $e$2, while $e$1 is the sum of four identical fragments) in the initial break up at 2.1 s. The regular fragments provided mass for the flares which started in time 5.8 s. Regular fragments, nevertheless, show nearly linear increase (in magnitude units) of brightness, while the light curve had a curved shape between 2.1 and 5.8 s. Two dust humps (marked $h$1, $h$2) from the eroding fragments were thus needed to obtain the observed profile.

The masses of the eroding fragments in the model were 18 kg and four times 13 kg. The regular fragments had masses 36, 21, and 20 kg. The mass of the largest fragment, 36 kg, is restricted by the observed deceleration (see Fig. 9). The total mass of regular fragments is determined by the energy radiated after time 5.8 s. The total mass and other properties of the eroding fragments have been set to explain the light curve profile between 2.1 and 5.8 s. The four $e$1 fragments had a quite low erosion coefficient of 0.04 $s^2$ $km^{-2}$ while $e$2 had 0.6 $s^2$ $km^{-2}$ and was therefore eroded out more quickly. In both cases the released dust particles had all the same mass of 1 gram. These relatively large bodies (almost cm-sized) ablated gradually and caused smooth humps on the light curve. The eroding fragments can be also interpreted as tightly packed agglomerates of pebbles, which are gradually lost from the edges.

Note that the derived masses depend on the assumed values of $\Gamma A$ and density. The masses would be lower for $\Gamma A = 0.7$ instead of 0.8 (24 kg instead of 36 kg for the largest fragment). However, the total mass would be not sufficient in that case to produce the flares after time 5.8 s. For $\Gamma A = 0.7$, the flares could be produced only assuming the luminous efficiency for dust 3.5% instead 2.5%. We prefer the value 2.5%, which was successfully used for other fireballs, but it is clear that mass values are model dependent. On the other hand, the location of the fragmentation points is not affected by the selection of parameters of the model.

The light curve shows many fluctuations between 2.1 – 5.8 s. It is not clear if they should be attributed to fragmentations or to some other effects like instabilities in evaporation and erosion, rotation of irregular fragments, fast chemical reactions, etc. We tried to model the two largest spikes at 2.3 and 3.2 s (heights 72 and 64 km) with fragmentation of a small fragment (marked 4 in Fig. 16). The spikes have amplitudes about one magnitude, durations of 0.1 s and nearly symmetrical shapes. To obtain short duration flares at such high altitudes, very small dust particles are needed to produce them: $10^{-12}$ and $10^{-11}$ kg, respectively (diameters of 8 and 18 micrometers). Symmetrical flares require gradual, not instantaneous, dust release. We used the formalism of eroding fragments with high erosion coefficients of 4 $s^2$ $km^{-2}$. Still, the symmetrical shapes of flares were not fully reproduced. In any case the masses needed to produce the flares were small, just 0.12 kg in both cases. If there were fragmentation events between 2.1 – 5.8 s, the involved mass was low in comparison with total meteoroid mass. We therefore did not attempt to reproduce all details of the light curve in this phase.

On the other hand, there is no doubt that the series of short flares, which started at time 5.8 s (height 40 km) were caused by fragmentation events. The duration of the flares was typically 0.03 s with the rise time of 0.01 s. We modeled these flares by instantaneous dust releases. The used dust particle masses were typically $10^{-5}$ kg (diameter 2 mm), in some cases extending to from $10^{-6}$ to $10^{-4}$ kg (~ 1 – 4 mm). The total dust masses were 3 – 4 kg in the brightest flares. The exact fragmentation sequence is, of course, unknown. In the model, the largest fragment 1 (mass 23 kg at that time) first broke into number of smaller fragments (marked 5 in Fig. 17, modeled as eight fragments of 2.4 kg each), which were responsible for the elevated brightness after the first couple of flares, but all disintegrated soon. After that, most light outside flares was produced by the dust from various eroding fragments, which emerged from all these disruptions. This 'dust' was, nevertheless, mostly composed from relatively large bodies, from 1 gram to 0.2 kg (diameters up to 5 cm). The erosion coefficients were in the range 0.05 – 0.1 $s^2$ $km^{-2}$. The situation cleared at time about 7 s, when the only surviving large fragment became the dominant source (marked 6 in Fig. 16). This is the fragment, which was observed on photograph to deviate from the original trajectory. Its properties were determined mostly from dynamics, i.e. the observed length (or height) as a function of time. The actual fragment trajectory was used in the model.

Table 3. Physical and atmospheric trajectory data on the Žďár nad Sázavou meteoroid

|  | **Beginning** | **Terminal** |
| --- | --- | --- |
| **Time (UT)** | 16:16:45.6 | 16:16:54.8 |
| **Velocity (km s$^{-1}$)** | 21.89 ± 0.02 | 4.8 ± 0.1 |
| **Height (km)** | 98.06 ± 0.02 | 24.71 ± 0.02 |
| **Slope (°)** | 26.035 ± 0.004 | 25.295 ± 0.020 |
| **Mass (kg)** | 150 ± 20 | ~1.3* |
| **Total length (km)/Duration (s)** | 170.5 / 9.16 | |
| **Maximum absolute magnitude** | -15.26 ± 0.10 at 36.8 km | |
| **P$_{max}$** | 2.7 MPa at 32 km | |
| **Radiated energy** | 1.4 x $10^9$ J (0.33 T TNT) | |
| **Fireball type/PE** | I / -4.39 | |

* the largest fragment

Explaining the dynamics was not easy. The velocity was lower than expected at the beginning of fragment formation (heights about 35 km) but was not decreasing very rapidly toward the end. The agreement with observations was finally reached under these assumptions: The final fragment did not originate from the largest initial fragment 1 but from the smaller fragment 3. After fragments 1 and 2 disintegrated (at times 5.78 – 5.81 s, heights 40.2 – 40.0 km), fragment 3, which was more decelerated at higher altitudes, became the leading fragment. This led to apparent decrease of fireball velocity. Fragment 3 broke only at later times, after 6.12 s (37.3 km), perhaps repeatedly. The most severe break-up at 6.18 s produced the brightest fireball flare at the height 36.8 km. Here, as we know from the geometry, the largest surviving fragment gained a lateral velocity of 0.2 km s$^{-1}$. To explain the dynamics, we assumed that at the same time, the forward velocity decreased by 0.3 km s$^{-1}$. The total velocity impulse of 0.36 km s$^{-1}$ was probably caused by momentum exchange with small fragments escaping asymmetrically. While the lateral component is quite certain, the backward component is less certain. The modeled velocity discontinuity is visible in Fig. 11.

After the main break-up, the modeled mass of the largest fragment was 3.5 kg. It was modeled to break once more at 6.52 s (34.2 km), where the mass decreased from 2.0 kg to 1.5 kg, producing one of the minor flares on the descending part of the light curve. The other flares at heights 34.0 – 31.4 km (until 7.0 s) were produced by disruptions of other fragments. The main fragment remained intact until the end of observation at height 24.7 km (~8.76 s, the last velocity measurement was at 8.72 s). To conform to the dynamics, the ablation coefficient was lowered to 0.001 s$^2$ km$^{-2}$ for this final part. The resulting terminal mass was 1.3 kg. The modeled brightness was somewhat higher than observed. It is possible that luminous efficiency decreased more quickly for velocities below 10 km/s than assumed. It also seems that radiometers underestimate fireball brightness near their limit of sensitivity.

Figure 17 shows the residuals of the dynamical fit, i.e. the differences between measured and computed lengths along trajectory as a function of time. The data before and after the geometrically observed break-up are shown separately since different trajectory was used in both cases. Since the fragmentation model does not consider gravitation, the expected trend due to gravitational acceleration along the trajectory ($d = \frac{1}{2} gt^2 \cos z$, where $g$ is gravity acceleration, $t$ is time, and $z$ is radiant zenith distance) was plotted. It is only a minor secondary effect. There is no systematic deviation of the data and most measurements are within 200 meters from the expected trend. Data from some stations show larger deviations due to unfavorable geometric conditions as discussed above.

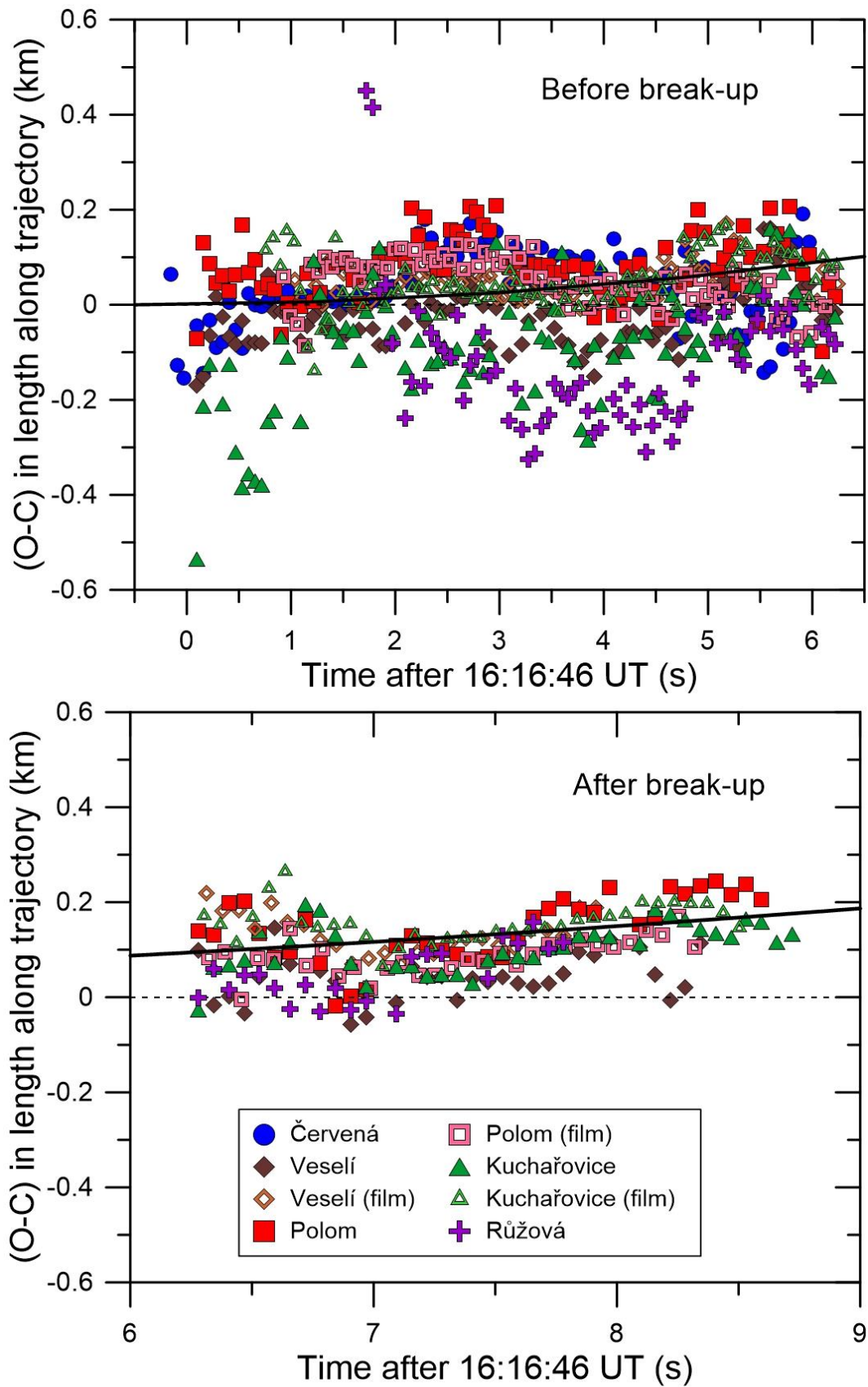

**Fig. 17.** Differences between measured and computed lengths along trajectory at individual cameras. Data before the main break-up (upper panel) and after that (lower panel) are shown separately because the trajectories were different. The solid lines show the expected trend due to gravity acceleration.

The fragmentation model is not able and is not intended to reveal all details of the fragmentation process, which was probably more complicated in reality. The aim is to describe the main phases of the fragmentation and to get insight into the strength and structure of the parent meteoroid. Meteoroids with different entry angles and velocities can be compared by comparing the dynamic pressures, at which they broke-up, and the amounts of mass lost. The dynamic pressure is computed as $p = \rho v^2$, where $\rho$ is atmospheric density and $v$ is meteoroid velocity. Figure 18 presents a comparison in this respect of Žďár nad Sázavou with two other ordinary chondrite meteorite falls, H5 chondrite Košice (Borovička et. al. 2013) and H6 chondrite Križevci (Borovička et. al. 2015). The mass of the largest surviving fragment is plotted as a function of increasing dynamic pressure. Borovička et. al. (2017) presented the same graph also for some large non-chondritic meteoroids. Note that the lowest plotted mass is not identical with the mass of the largest expected meteorite. Both Košice and Križevci fragmented further at the end of the trajectory, when the dynamic pressure was already decreasing (because of deceleration).

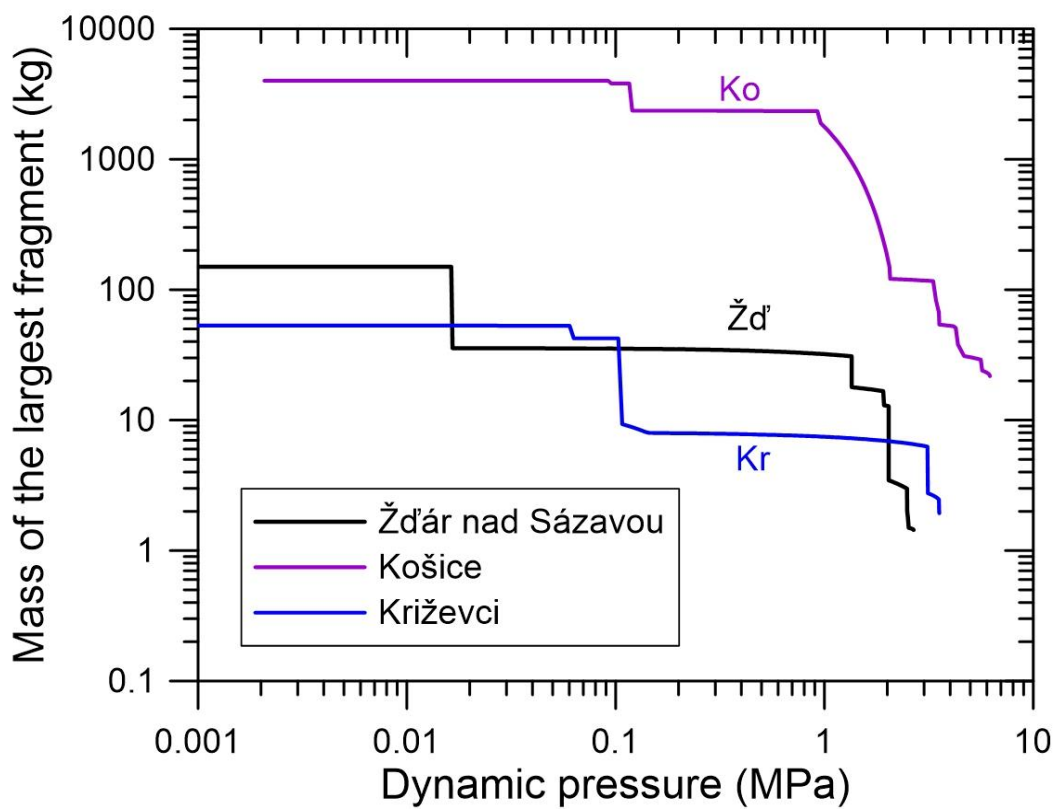

**Fig. 18.** Mass of the larges surviving fragment (according to the fragmentation model) as a function of increasing dynamic pressure for three ordinary chondrite meteorite falls.

All three meteoroids are characterized by two-stage fragmentation. After the initial significant break-up, where about a half (in case of Košice) or even more mass was lost, no further fragmentation occurred until the dynamic pressure increased by an order of magnitude (to about 1 MPa) or even more. Then a series of fragmentation followed. The second phase occurred under similar pressures in all three cases: 1 – 5 MPa in Košice, about 3 MPa in Križevci, and 1.4 – 2.5 MPa in Žďár nad Sázavou (where the largest remaining fragment survived intact maximum pressure of 2.7 MPa). On the other hand, the initial break-up of

Žďár nad Sázavou occurred already at 0.016 MPa, a pressure almost an order of magnitude lower than in other two cases. For both Košice and Križevci, the initial break-up occurred at about 0.1 MPa. Žďár nad Sázavou was therefore much weaker meteoroid and soon disintegrated into primary fragments. These fragments, nevertheless, had similar strengths as primary fragments of other two meteoroids.

## Other data

*Sound*

Along with the photographic imaging system and radiometer the film cameras are equipped by a simple microphone. Six minutes of sound is recorded to the hard disk after bright events detected by radiometer. In most cases, no fireball signature is detected but the sonic boom from Žďár nad Sázavou fireball was clearly detected at the closest station Svratouch at 16:18:47.3 UT, i.e. 113 seconds after the end of the fireball (Fig. 19). Possible weak sound arrivals are present on filtered record also between 16:18:49–51 UT.

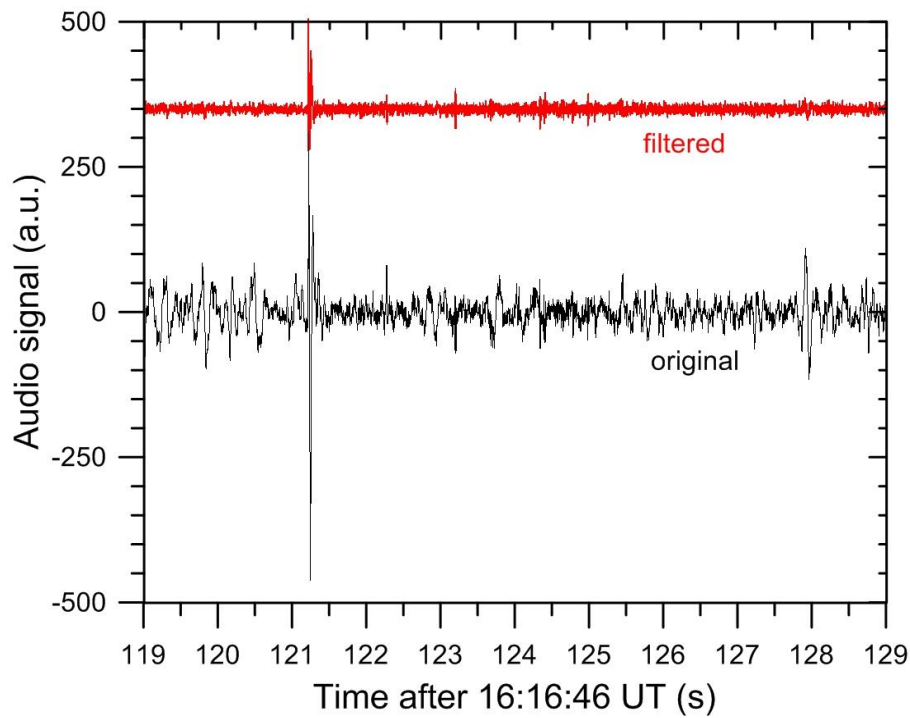

**Fig. 19.** Audio record at station Svratouch. The upper curve (offset by 350 units for clarity) was obtained after application of high pass filter with cut-off frequency 100 Hz to the original data using the Audacity software.

The closest point of fireball trajectory to station Svratouch lies at height of 24.6 km, just behind the observed fireball end. The largest fragment was expected to pass this point at time 16:16:54.85 UT, when it had still supersonic velocity of 4.5 km s$^{-1}$ and a mass of 1.3 kg. The distance to Svratouch is 34.36 km. The mean speed of sound between that height and the altitude of Svratouch was 302 m s$^{-1}$, so that the earliest arrival of sonic boom at Svratouch can be expected at 16:18:48.6 UT. This crude estimate could be further improved by ray tracing taking into account wind field, nevertheless, it is already very close (within 1.3 s) to the actual time. We can therefore conclude that the observed sonic boom was caused by cylindrical blast wave originating from the closest point on the trajectory. It is noted that relatively small fragment moving at relatively low speed caused detectable signal though the instrument is not particularly sensitive. There were also detections of sonic waves on seismic stations (P. Kalenda, private comm.).

*Video*

A casual video record of the fireball was obtained by Mr. Jiří Hlávka with his dashboard camera from moving car during a drive from Trnava in the direction to Senica, Slovakia. The distance to the fireball was of the order of 150 – 200 km. The video has a resolution of 1280 × 720 pixels and 30 frames per second, though many frames are missing and instead of them, repeating frames are inserted. The video was not calibrated astrometrically, nevertheless, two bright flares separated by 0.4 s can be easily seen and identified with two flares at times 5.8 s and 6.2 s in the radiometric light curve (here we use the same time scale as in Figs. 16 and 17). Video frames therefore provide snapshots of fireball appearance at various times. Though the resolution is low, two or more fragments can be seen in a number of frames. We provide a sample of fireball images in Fig. 20.

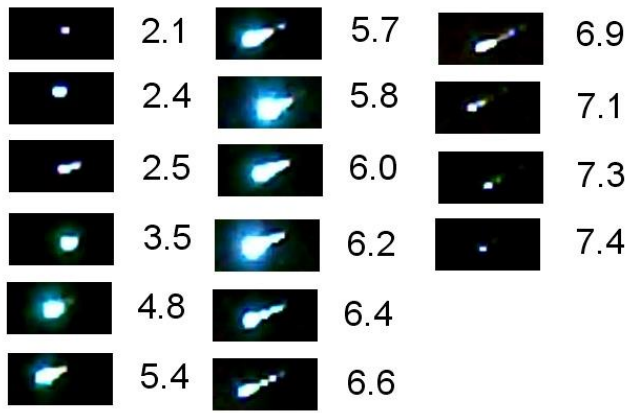

**Fig. 20.** Snapshots of the fireball extracted from a casual dashboard video obtained by Mr. Jiří Hlávka. The time in seconds after 16:16:46 UT is shown to the right of each image.

In general, the video confirms our fragmentation analysis. A fragment appears at time 2.5 s, which must be a group of tiny particles released at the earliest fragmentation at the height of 74 km. Atmospheric drag is insufficient at such a height to separate larger fragments. They become separated about 3 seconds later. More fragments are visible after the main flares. After the time 7.4 s, only one dominating fragment remains visible. Terminal fragmentation was recorded also by the Slovak AMOS system (J. Tóth, private comm.).

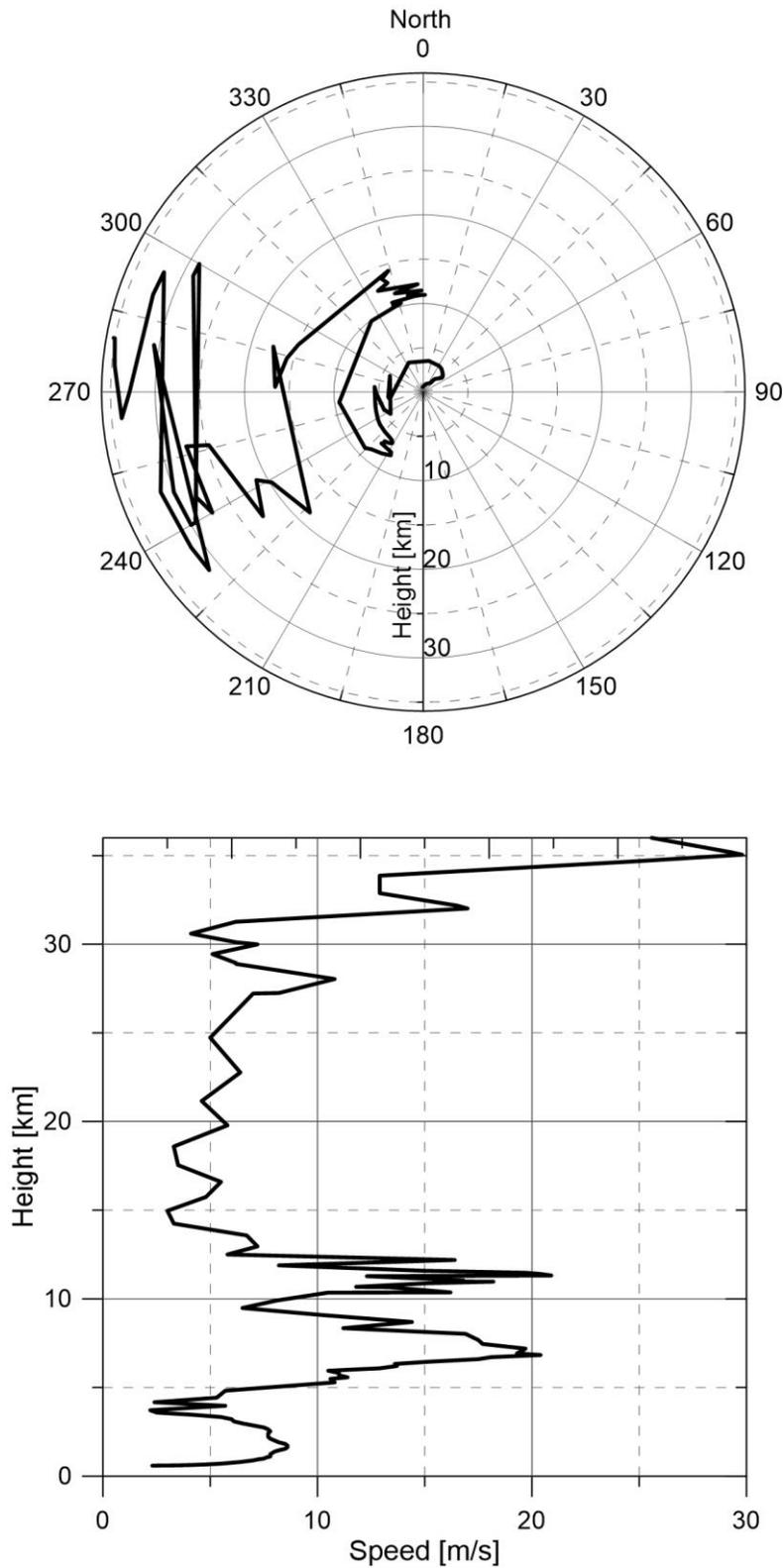

**Fig. 21.** Vertical wind profile used for dark flight computation. The source data are airplane measured winds at heights 4 – 13 km (courtesy J. Kráčmar, Air Navigation Services of the Czech Republic), the forecast for 16 UT by the numerical model ALADIN (the run from 12 UT, courtesy R. Brožková, Czech Hydrometeorological Institute), and the radiosonde measurements from Prague on 12 UT, which were used above 27 km (data downloaded from University of Wyoming site at http://weather.uwyo.edu/upperair/sounding.html). The upper polar plot shows wind direction, the lower plot shows wind speed.

## Expected meteorite distribution

The results of the fragmentation model were used to estimate the number, masses, and locations of meteorites. The model tracked all fragments until their velocity decreased to 2.5 km s$^{-1}$. At that point, the ablation effectively ceases and the fragments continue dark flight to the ground with unchanged mass. One dominant meteorite with the mass of 1.3 kg is predicted to exist (in case of no splitting during dark flight). The second largest meteorite may be in the range 100 – 200 grams. The model predicts ~250 meteorites in the mass range 10 – 200 g (total mass 6 kg) and 3000 meteorites in the range 1 – 10 g (total mass 7 kg). These numbers must be, however, considered as upper limits. The model neglects any further fragmentation of bodies released formally as "dust". Such secondary fragmentations surely exist but in unknown proportion. Also large fragments can undergo unnoticed fragmentations, even during the dark flight. This is evidenced by fresh meteorites with fusion crust missing or being extremely thin at a significant part of the surface as for Žďár M1 meteorite (see Fig. 23) or e.g. in case of the Morávka meteorite fall (Borovička and Kalenda 2003).

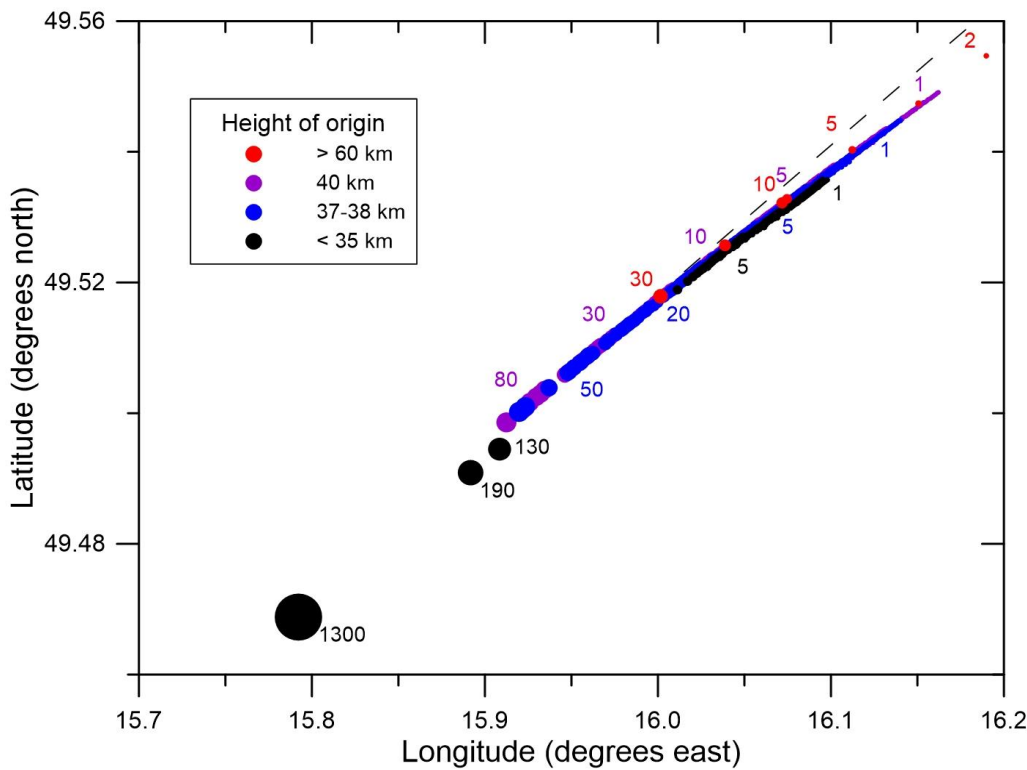

**Fig. 22.** Predicted coordinates of meteorite falls. The symbol size is proportional to meteorite size. Labels indicate masses of selected meteorites in grams. Different colors correspond to origin of meteorites in different heights. The dashed line is ground projection of fireball end. Note that scales on x- and y-axes are different.

The major issue for correct computing of meteorite landing positions was the high atmospheric wind profile. Radiosonde measurements in Prague showed a major wind change between December 9, 12 UT, and December 10, 0 UT. This resulted in the shift of the predicted meteorite position by 2 km southwards for the largest meteorite and by almost 6 km for one-gram meteorites. We therefore asked R. Brožková from the Czech Hydrometeorological Institute to provide us with the forecast for 16 UT from the 12 UT run of the numerical meteorological model ALADIN. Because the main change of wind direction occurred at altitudes 8-10 km, we also contacted the airplane traffic control in Prague and asked them if they have data from airplanes. They compiled data based on radar detected motions of airplanes compared with on board measured velocity vectors from airplanes flying within 15 minutes and 50 km from the fireball end. Both ALADIN and air control showed that the wind change did not yet occurred in the region of interest at the time of the fireball. From all available data we prepared the most probable wind profile (showed in Fig. 21) and used it for the dark flight computations.

Figures 22 and 23 shows the predicted meteorite locations. By chance, the strongest wind at height around 8 km blew from south-west, nearly in opposite direction to fireball flight. In result, fragments of all masses are aligned along a line, which is very long. The distance between one-gram fragments originating in the fragmentation at height 40 km and 100 g fragments is 20 km. The largest meteorite is predicted to lie further 7 km ahead. The three larger meteorites deviate to the south because the observed trajectory change at the height of 36 km. All other fragments were assumed to follow the main trajectory. In reality, small fragments could also gain side velocities and further lateral spread can result from irregular fragment shapes. The strewn field is therefore expected to be, as usually, several kilometers wide. The plotted points just define the central line.

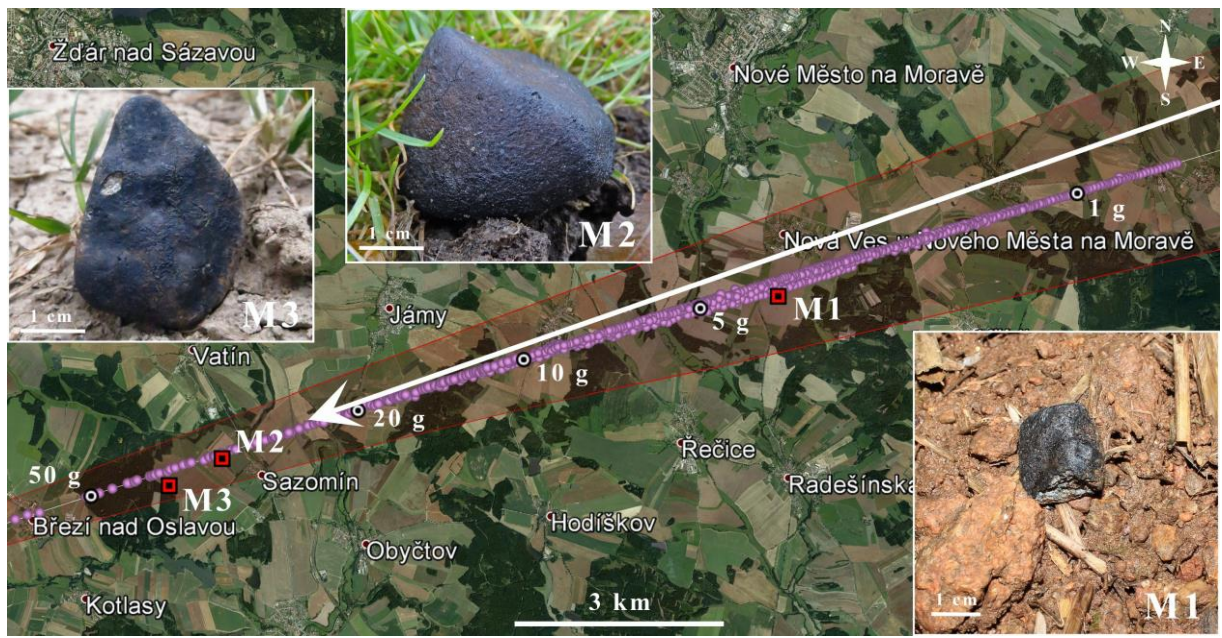

Fig. 23. Map of the predicted impact area for small pieces (1-50g) with marked positions of all three recovered meteorites and their images in finding positions (Source of the background image: Google Earth).

For the sake of completeness, we also modeled meteorite falls from the earliest fragmentation at the height of 74 km. A release of fragments treated as a dust was added to the model (marked $d$1 in Fig. 16), although they were not necessary for the explanation of the light curve. In addition, one modeled meteorite originated from the disruption of fragment 4 at the height of 64 km. As it can be seen from Fig. 22 (red points), these high born meteorites, if they exist, follow the same line as other meteorites, only are shifted along the line backwards (to the east) in respect of other meteorites of the same mass. The map containing the central line for the highest number of meteorites covering fragments from 1-50g, probable impact area, predicted sizes of meteorites and all three recovered pieces with images in finding positions is given in Fig. 23. Whole modelled impact area covering all masses from the smallest meteorites to the main piece is in Fig. S1 of supporting information of the Appendix S1.

## Search for meteorites and meteorites recoveries

The impact area is located in the low populated Bohemian-Moravian Highlands south of the town Žďár nad Sázavou. The area is covered mostly by agricultural land, i.e. plowed, sown, and stubble fields, grasslands, forests, ponds, but also one dam and several small villages. Terrain of this area is not very favorable for search for meteorites. Moreover, due to the winter season snow and ice affected the searching in the first 4 months after the meteorite fall, and after that intensive spring field works and growing vegetation quickly changed the surface. In combination with vastness of the area, these circumstances made the systematic search very difficult and with the restricted capacities, we were able to concentrate only on the best searchable parts.

The first visit and search in the impact area took place already 18 hours after the fall, when a small group of us made a rough reconnaissance of the largest pieces area and shortly searched in the expected area of the main piece. However, the proper wind profile (discussed in the previous section) was unknown at that time so we were a bit southwards.

Since the predicted impact area is more than 30 km long (1 g to 1 kg fragments), we decided soon to ask local people, amateur astronomers, and enthusiastic individuals for help with meteorite searching. On December 18, 2014, we informed inhabitants from the villages lying in the most probable parts of the impact area by posters where we described the possibility of casual meteorite find near to their houses. Only one inhabitant reported suspicious stone, which turned to be not a meteorite after our inspection. Important help came from three different subjects, which offered co-operation and participated in searching: Astronomical Society of Jihlava, Interplanetary Matter Society, and Czech journal Vesmír (The Universe). This co-operation resulted in 15 searching days and the first recovered meteorite (M1, see Fig. 23). This meteorite was found in the area of 1-10 g meteorites originated in two largest flares by T. Holenda, member of the group of amateur astronomers from the Astronomical Society of Jihlava and Interplanetary Matter Society, near the end of the first searching day on December 20, 2014. This small freshly looking fragment weighing 5.9 grams was partly broken. The first snow occurred still in the evening after the find and the area was quickly covered by thick layer of snow, which made further search impossible until January 10, 2015, when snow quickly melted. It enabled to continue in searching and resulted in the find of the second meteorite (M2, Fig. 23). It was recovered by T. Henych, a member of group of five people from our department on January 12, 2015 some 8.3 km apart from the M1 towards to larger masses. This meteorite weighing 39.3 g (after 2 days of drying, the recovery mass was 41.1 g – see details in Kalašová et al., 2020) was completely covered by a fusion crust and similarly like M1 looked fresh. Finally, the third meteorite of similar size and appearance like M2, was recovered by private collector Z. Tesařík on May 2, 2015 some 850 m ahead towards larger masses from the site of M2 (M3, Fig. 23). Its weight is 41.7 g and this piece is noticeably more affected by weathering. M1 and M3 were recovered on sown fields, M2 on a meadow. Basic data on the recovered meteorites are collected in Table 4 and finding locations in Fig. 23.

Altogether half a year of discontinuous searching resulted in three meteorites known to us. We searched almost 3.6 km$^2$ out of 6 km$^2$ (some fields and forests were excluded due to unfavorable terrain) in 42 days between December 10, 2014 and June 11, 2015. Since we searched for fresh meteorites, we used only our eyes and walked half a meter to few meters apart depending on the clarity of the terrain. No metal detector or any magnetic equipment was used for searching. All three pieces were found exactly in the predicted area for a given mass (Fig. 23). First two meteorites are deposited at the Astronomical Institute of the CAS, the third one is owned by the finder.

These successful recoveries of three Žďár nad Sázavou meteorites, which were classified as a L3 ordinary chondrites of subtype 3.9, shock stage S2 and weathering grade W0 (Kalašová et al., 2020), perfectly confirmed our analyses and prediction of meteorites position.

Table 4. Details of the recovered meteorites Žďár nad Sázavou

| Met. No. | Date of find | Weight (g) | Coordinates Longitude E | Latitude N | Finder |
|---|---|---|---|---|---|
| M1 | 20. 12. 2014 | 5.92 | 16.08326 | 49.53205 | T. Holenda |
| M2 | 12. 1. 2015 | 39.25 | 15.97327 | 49.51102 | T. Henych |
| M3 | 2. 5. 2015 | 41.70 | 15.96280 | 49.50751 | Z. Tesařík |

## Conclusions

The instrumentally observed meteorite fall Žďár nad Sázavou occurred in the Czech Republic on 9 December 2014 at 16:17 UT. The original meteoroid with an initial mass of 150 kg and diameter approximately 45 cm entered the atmosphere with a speed of 21.89 km s$^{-1}$. It began its luminous flight at an altitude of 98.06 km. In maximum, it reached -15.26 absolute magnitude and terminated after 9.16 s and 170.5 km long flight at an altitude of 24.71 km with a speed of 4.8 km s$^{-1}$. The average slope of the atmospheric trajectory to the Earth's surface was only 25.66°. The total radiated energy was $1.4\times10^9$ J which corresponds to 0.33 T TNT. From the detailed light curve containing several well-defined distinct flares, we determined that the meteoroid repeatedly fragmented during its flight. First fragmentation, when the initial meteoroid disintegrated into primary fragments, occurred already at a height of about 74 km, under surprisingly low dynamic pressure of 0.016 MPa. It means that original Žďár meteoroid was a relatively fragile meteoroid, very probably due to previous collisions in interplanetary space. These fragments, nevertheless, had much higher strengths and fragmented at lower heights under larger pressures. The largest remaining fragment survived intact maximum pressure of 2.7 MPa, which is comparable with similar cases as Košice (Borovička at al., 2013) or Križevci (Borovička at al., 2015).

Before its collision with Earth, the initial meteoroid orbited the Sun on a moderately eccentric orbit with perihelion near Venus orbit, aphelion in the outer main belt, and low inclination. It has an orbit of obviously asteroidal origin with the Tisserand parameter relatively to Jupiter 3.42 and orbital period of 3.03 years. Two small near-Earth asteroids 2011 WU74 and 2011 WV74 have orbits of the same character, though with little bit different orientation and higher inclinations (5.9° and 7.2°, respectively).

Based on our analysis, so far three small meteorites classified as L3 ordinary chondrites totaling 87 g have been found almost exactly in the locations predicted for a given mass. Žďár nad Sázavou is the fourth meteorite with precisely known orbit after Příbram in 1959 (Ceplecha 1961), Benešov in 1991 (Spurný et al., 2014), and Morávka in 2000 (Borovička et al., 2003) found in the Czech Republic. All these exceptional cases were completely analyzed by the team from the Astronomical Institute of the CAS in Ondřejov.

Main exceptionality of this case consists in detailed analysis of high number of high-resolution instrumental records (both photographic and radiometric) taken by the automated fireball cameras (both digital and analog) in the Czech part of the European fireball network. Thanks to our experience how to analyze such unusually large amount of high quality data from dedicated instruments, we were able to obtain detailed, reliable and precise results concerning atmospheric trajectory, photometry, dynamics, and heliocentric orbit of the Žďár bolide. On top of that, we described in detail its fragmentation scenario and predicted impact area for all sizes of meteorites originated in several break-ups and terminal point as well. The reality, i.e. recovered meteorites, exactly confirmed our predictions because all meteorites were found in the predicted areas and their masses well correspond to the predicted ones. This is the best proof of correct interpretation of our observations, and correctness of our analyses and methods as well. The Žďár nad Sázavou bolide and meteorite fall was the first, which was recorded by the newly modernized Czech part of the EN and validated correctness of this modernization of our instrumental facilities and their background as well.  This fact was subsequently confirmed by another similarly recorded and analyzed meteorite falls as Stubenberg (March 6, 2016 in Germany, Spurný et al., 2016, Bischoff et al., 2017), Hradec Králové (May 17, 2016 in Czech Republic, Spurný and Haloda 2017), and Renchen (July 10, 2018 in Germany, Spurný et al., 2018, Bischoff et al., 2019), which were recovered on the grounds of our records and precise description and prediction as well.

## Acknowledgements

This work was supported by Praemium Academiae of the CAS, grant GA ČR 16-00761S and institutional project RVO:6798581. We are very much indebted to R. Brožková for the ALADIN wind profile model, the staff members of the Air Navigation Services of the Czech Republic, namely J. Kráčmár and J. Frei,  for the stratospheric wind data from the airplanes flying over the impact area, I. Macourková and J. Zákopčaník for mediation of these valuable data, J. Hlávka who provided us with his dashboard camera video record, and especially all three meteorite finders T. Holenda, T. Henych, and Z. Tesařík as well as members of Astronomical Society of Jihlava represented by M. Podařil, Interplanetary Matter Society represented by J. Kondziolka, and Czech journal Vesmír (The Universe) represented by the main organizer M. Janáč for their important help in searching. In this respect we also thank to members of the Interplanetary Matter Department namely (in alphabetical order) D. Čapek, K. Hornoch, Z. Karas, J. Kašpárek, J. Keclíková, P. Koten, L. Kotková, P. Kušnirák, J. Starý, P. Scheirich, L. Smolíková,, R. Štork, V. Vojáček, and H. Zichová as well as external searchers A. Martínková, A. Spurná, P. Spurný, Jr., and P. Trepka for their searching effort. Our special thanks go to J. Keclíková who measured all photographic images, and to J. Boček and J. Zeman who helped us with deployment of the new digital cameras manufactured by cooperating company Space Devices, Ltd. on the stations of the Czech part of the EN.

## Supporting information

Additional supporting information may be found in the online version of this article.
**Appendix S1**. Station coordinates, fireball dynamics and photometric data (file Supplementary_data.xlsx), and impact area details for the Žďár nad Sázavou fireball (file FigureS1.jpg).